\documentclass[12pt,a4paper]{article}
 
\pdfoutput=1 
\usepackage{graphicx}
\usepackage{jcappub}
\usepackage{amsmath,amsthm,latexsym,amssymb,amsfonts,epsfig}
\usepackage{hyperref}
\usepackage{psfrag}
\usepackage[cp1250]{inputenc}
\usepackage[usenames,dvipsnames]{xcolor}
\usepackage{float}
\usepackage{tabularx}

\newcommand{\be}{\begin{equation}}
\newcommand{\ee}{\end{equation}}
\newcommand{\bea}{\begin{eqnarray}}
\newcommand{\eea}{\end{eqnarray}}
\numberwithin{equation}{section}
\usepackage{physics}
\numberwithin{equation}{section}

\title{Sourced fluctuations in generic slow contraction}

\author[a,b]{Micha{\l} Artymowski,}
\author[a]{Ido Ben-Dayan,}
\author[a]{Udaykrishna Thattarampilly}
\affiliation[a]{Physics Department, Ariel University, Ariel 40700, Israel}
\affiliation[b]{Cardinal Stefan Wyszynski University, College of Science, Department of Mathematics and Natural Sciences, Dewajtis 5, 01-815 Warsaw, Poland}
\emailAdd{michal.artymowski@fuw.edu.pl}
\emailAdd{ido.bendayan@gmail.com}
\emailAdd{uday7adat@gmail.com}

\abstract{We introduce a set of generic conditions for the slow contracting Universe and for a narrowed-down category of  models called fast-roll models. We present general conditions for superhorizon freeze-out of scalar and tensor perturbations and show that any fast-roll model satisfies them, as in the case of inflation. We are interested in the "Sourced Bounce" scenario, where perturbations are sourced by a $U(1)$ gauge field coupled to a bouncer scalar field. The requirement of a slightly red tilted scalar spectrum greatly restricts the allowed couplings between the scalar and the gauge field. We show that a viable slightly red scalar spectrum is achievable. However, within the fast-roll approximation the tensor-to-scalar ratio is in general $r\simeq 1/9$, inconsistent with current observations. We demonstrate the general result with an explicit example we dub "Intermediate contraction". We prove that small modifications in fast-roll that do not alter the Green's functions,  do not result in $r<0.06$ consistent with the data for more than an e-fold. Hence, a successful "Sourced Bounce" requires a different source or a significant deviation from fast-roll.}
 

\begin{document}

\maketitle

\section{Introduction}

Current theories and models of our Universe are founded upon a myriad of astrophysical observations \cite{Aghanim:2018eyx,Ade:2018gkx,Simard:2017xtw,Louis:2016ahn,Akrami:2018odb,Inserra:2020uki,Hinshaw:2012aka,Aubourg:2014yra,Salazar-Albornoz:2016psd}. These observations imply that the Universe is at large homogeneous and isotropic with inhomogeneities displaying a red tilted (nearly scale invariant) scalar power spectrum. There are two well-known theoretical frameworks, which successfully provide an explanation for the observed homogeneity and isotropy of the Universe with successful models predicting a red tilted spectrum. The first one is cosmic inflation \cite{Martin:2013tda}, which is a hypothetical era of accelerated expansion in the early Universe. The accelerated expansion can be obtained by assuming that the Universe is filled with a scalar field called the Inflaton with an equation of state $w = p/\rho < -1/3$ caused by a flat positive scalar potential. Rapid, quasi-exponential growth of the scale factor homogenizes the Universe. The quantum fluctuations of the Inflaton quite naturally generate slightly red tilted spectra of primordial gravitational waves (GW) and curvature perturbations. Despite its success and consistency with observational data, the theory of inflation has a few problems, including possible fine-tuning of initial conditions \cite{Ijjas:2013vea}, simple models being ruled out \cite{Ade:2018gkx}, and the fact that it does not resolve the big bang singularity \cite{Borde:1996pt,Borde:2001nh}. 

A popular alternative to inflation is its complete opposite - instead of a flat, positive potential and accelerated expansion of the Universe, one can consider a slowly contracting Universe driven by a bouncer field, which is a scalar field with a steep negative potential. A contracting Universe tends to suffer from the growth of anisotropic stress, which redshifts like $a^{-6}$. A bouncer field that has an equation of state $w>1$ means that its energy density $\rho$ redshifts faster than $a^{-6}$. Hence, the anisotropic stress remains subdominant. The fast redshift of a bouncer field homogenizes the Universe, providing the solution to the flatness and horizon problems. Thus, slow contracting models are free of the BKL instability suffered by more conservative models like the matter bounce \cite{Wands:1998yp,Brandenberger:2012zb,BATTEFELD20151}. The same mechanism of quantum fluctuations is at work here as well: Curvature perturbations generated during the slow contracting phase of the Universe give rise to large scale structure. The slow contraction is followed by the bounce of the scale factor at finite value, and therefore it provides not only the solution to the problems of isotropy and large scale structure of the Universe, but also sidesteps the problem of a primordial singularity \cite{BATTEFELD20151,Cai:2008qw,Lehners:2010fy,Cai:2012va} without the need to modify gravity, or a full-fledged quantum gravity theory.

The best known example of slowly contracting Universe is the ekpyrotic model, defined by a power-law evolution of the scale factor. A recurring issue of ekpyrotic models driven by a scalar field is blue tilted power spectra of curvature and GW perturbations \cite{Lehners:2010fy,BATTEFELD20151}. This unwelcome feature can be dealt with by using an additional degree of freedom as a curvaton field, which is responsible for generating the primordial inhomogeneities and relaxes constraints on the ekpyrotic field. The curvaton is usually a scalar field, however such models do not alter the tensor spectrum significantly \cite{Creminelli:2014wna,Geshnizjani:2014bya}. 

The idea that sourced fluctuations could considerably modify the predicted tensor spectrum in bouncing models was first suggested in \cite{r1,r3}, with more work along these lines in \cite{Chowdhury:2016aet,Chowdhury:2015cma,Chowdhury:2018blx,Gasperini:2017fqw,Ito:2016fqp}. Considering a gauge field coupled to the bouncer field, rather than another scalar field is by no means more "fine-tuned" or a more "complicated" model. On the other hand it allows getting a nearly scale invariant scalar and gravitational spectra due to fluctuations sourced by the gauge field, undermining the idea that measuring the tensor-to-scalar ratio $r$ provides a "proof of inflation". Furthermore, the tensor spectrum may be slightly blue and chiral, in contrast with most of the inflationary models. \footnote{Let us note that there are quite a few inflationary models, which predict slightly blue tensor spectrum \cite{Lin:2015nda, Wang:2014kqa} and which may be motivated by the presence of gauge fields \cite{Mukohyama:2014gba}. These models are certainly not the standard canonical single field models within GR.} If such a model is to be realized in nature, one may not only measure $r$, but also detect a stochastic gravitational waves background by Laser Interferometers, as well as the chirality of the spectrum. \cite{LIGOScientific:2019vkc,Abbott:2016blz,AmaroSeoane:2012km,PhysRevD.81.123529} 

When considering sourced fluctuations, the first and foremost consideration is whether the analysis is consistent and the backreaction of the sourced fluctuations on the background evolution is under control. In past works, the source was a $U(1)$ gauge field. For that type of source, one had to avoid an IR divergence of the sourced fluctuations.  The sourced fluctuations generated a slightly blue tilted tensor and scalar spectrum for an ekpyrotic Universe that underwent a power law contraction within the backreaction limitations \cite{r3,r1}. However, the tensor-to-scalar ratio $r$ was found to be way over acceptable bounds from observations \cite{r1}.

We would like to devise a successful model, i.e. a slow contracting Universe with sourced fluctuations free of backreaction that reproduces the observed scalar spectrum and gives a tensor to scalar ratio within the allowed bound. Hence, a successful model requires either a different source, a different coupling between the source and the bouncer field, or a different background evolution. In this work, we discuss the two latter possibilities. We consider more general couplings between the gauge field and the bouncer field, as well as different background evolution that still fulfills so-called "fast-roll" conditions.

Given a $U(1)$ gauge field as the source, we show that the nearly scale invariant scalar spectrum greatly restricts the allowed couplings between the bouncer and the gauge field\footnote{A linguistic comment: The observed scalar spectrum is of course nearly scale invariant with $n_S\simeq 0.96$. However, throughout the manuscript we will drop the word "nearly" to reduce clutter. We will show that one can get a nearly scale invariant scalar spectrum in the slow contracting models as the data suggests.}. We then show that $r\simeq 1/9$ is an expected result in a broad family of slowly contracting models, which satisfy the fast-roll approximation. Furthermore, we show that deviations from this approximation that could lead to acceptable lower bound on $r$ can only be relevant for less than one e-fold. The main obstacle for a successful model is the narrow set of couplings resulting in a nearly scale invariant scalar spectrum, which then constrains $r$ to be $r\simeq 1/9$. We demonstrate this general result  explicitly in a scenario we dub "Intermediate contraction". Thus, any slowly contracting Universe, which satisfies the fast-roll approximation and has fluctuations sourced by a $U(1)$ gauge field generating power spectra is inconsistent with the data. A successful model must either deviate from the fast-roll approximation, or have a different source than the $U(1)$ gauge field discussed so far.

The backreaction issue can be addressed in two ways. First, by limiting the allowed parameter range as before, \cite{r3,r1}. Second, if the slow contraction phase is of finite duration. For example, if the potential is negative only at finite field values, while it is positive in others. In such a case, the Universe always contracts, but when the potential is positive, one has accelerated contraction and sourced fluctuations are not produced in a significant amount. This happens, for instance, in the  Intermediate contraction scenario. In any case, we would like to stress that even disregarding the backreaction does not lead to a viable $r$. 

The structure of this paper is as follows. In Sec. \ref{sec:slow} we introduce general conditions for the slow contraction together with the fast-roll approximation. In Sec. \ref{sec:pertun} we present the general conditions for the freeze-out of scalar and tensor perturbations in a contracting Universe and present generic superhorizon solutions of Green's functions. In Sec. \ref{sec:gauge} we present the model with $U(1)$ gauge field, and discuss the implications of a scale invariant spectrum. In Sec. \ref{sec:r} we show that sourced tensor-to-scalar ratio cannot be consistent with the data for any fast-roll slowly contracting Universe. We give an explicit example of our result, that we dub Intermediate contraction in Sec. \ref{sec:ic}. Finally, we conclude in Sec. \ref{sec:concl}. We relegate several technical derivations to appendices.


\section{Slow contraction and fast roll approximation} \label{sec:slow}
Bouncing cosmology is an alternative explanation for the observed properties of the Universe - a homogeneous, flat, and isotropic background, and nearly scale-invariant primordial fluctuations. Bouncing models solve the problems of the Hot Big Bang such as the horizon problem and flatness problem without invoking inflation. Another virtue of many bouncing models as the ones discussed here is that they are geodesically complete, i.e. singularity free, unlike the inflationary Universe\footnote{One should remember that similar tools employed to generate the bounce in slow contracting models (such as modification of gravity or non-canonical kinetic terms) can also be used to get rid of the primordial singularity before inflation (see e.g. Ref. \cite{Wan:2015hya,Agullo:2013ai}). The key difference between these approaches is the fact that for inflation, the bounce is just a "patch" to get rid of the singularity and not inherent to inflation, or necessarily having observational signatures, while for the slow contracting scenario, a bounce is a necessity. We strive for a more complete picture where a single model both discards the singularity and generates the observed spectrum.}. Let us briefly discuss how the classical problems mentioned above get solved by a contracting model. The technical details can be reviewed for instance in \cite{Lehners:2010fy,BATTEFELD20151}. 

The horizon problem is resolved if far separated regions of the Universe today were in causal contact during a previous contraction phase. For a slowly contracting Universe the comoving horizon is decreasing with contraction. Hence, given that Universe was contracting for sufficient number of e-folds far separated regions today were indeed in causal contact with each other during the early phases of contraction. Curvature grows as $a^{-2}$ during the slow contraction phase, this growth is subdominant to the growth of energy density of the bouncer field $(w>1)$, hence the curvature contribution can be made as small as desired during the slow contraction. By requiring that the bounce is sufficiently symmetric, the curvature contribution at the end of the contracting phase is the same as that at the beginning of the expansion phase. To summarize, the flatness and horizon problems are resolved by requiring that the contracting phase lasts for sufficient number of e-folds ($\simeq60$). 

Let us consider a slowly contracting FLRW Universe with a scale factor $a(t)$ and Hubble parameter $H(t) \equiv \frac{\dot{a}}{a}$, where the slow contraction (SC) is defined by two conditions
\begin{equation}
H < 0 \qquad \text{and}\qquad \epsilon \equiv -\frac{\dot{H}}{H^2} = - \frac{H_N}{H} > 3 \, , \label{eq:slowcontrcondition}
\end{equation}
where $\dot{H} = \frac{dH}{dt}$, $H_N = \frac{dH}{dN}$ and $N=\log a$ is the number of e-folds counted from the end of slow contraction. Furthermore, let us assume that the slow contraction happens between some $t=t_i<0$ and $t=t_f<0$, where $t_i<t_f$. $t_i$ is the moment of the beginning of the slow contraction and in principle it can take an infinite value, as in the case of ekpyrosis. $t_f$ is determined by an additional mechanism that leads to the bounce. Common examples are non-canonical kinetic terms or a modification of gravity. We will assume that the predictions derived in this work are not modified by the specific bouncing mechanism. 

For small, perturbative parameters, let us consider $\tilde{\epsilon} = \epsilon^{-1}$. Since $\epsilon>3$ one finds $0 < \tilde{\epsilon} < 1/3$. Throughout at least 60 e-folds of slow contraction one finds $\Delta\tilde{\epsilon} = |\tilde{\epsilon}(t_i) - \tilde{\epsilon}(t_f)| \leq 1/3$. Thus, it is safe to conclude that in order to keep $\tilde{\epsilon}$ small throughout at least 60 e-folds, barring cancellations, one requires
\begin{equation}
\left|\tilde{\epsilon}_N\right| \ll 1 \, , \quad \left| \tilde{\epsilon}_{NN}\right| \ll 1 \qquad \Rightarrow \qquad \left| \frac{\epsilon_N}{\epsilon} \right| \ll \sqrt{\epsilon} \, , \quad \left| \frac{\epsilon_{NN}}{\epsilon} \right| \ll \epsilon \, . \label{eq:slowcontrgen}
\end{equation}
The most generic form of the slow contraction brings rather limited information on the evolution of both background and inhomogeneities. Thus, just like the case of inflation, where the $\epsilon<1$ condition is narrowed down to the slow-roll approximation with $\epsilon \ll1$, $\epsilon_N/\epsilon \ll 1$, let us introduce the fast-roll approximation defined by
\begin{equation}
\epsilon \gg 1 \, ,\qquad \left|\frac{\epsilon_N}{\epsilon}\right| \ll 1 \, , \qquad \left|\frac{\epsilon_{NN}}{\epsilon}\right| \ll \epsilon \, . \label{eq:FRapprox}
\end{equation}
Notice that \eqref{eq:FRapprox} is more restrictive than \eqref{eq:slowcontrgen}. Two examples of models of contraction that fully satisfy the fast-roll approximation are: 
\begin{itemize}
\item[a)] Power-law contraction: This exact solution of Friedmann equations is generated by a scalar field with a canonical kinetic term and the background evolution of the form
\begin{equation}
a \propto (-t)^f \, , \  V = - f(1-3f) \, e^{\pm \sqrt{\frac{2}{f}}\, (\phi-\phi_0)} \, , \  \phi = \phi_0 \mp \sqrt{2f}\log (-t) \, , \  \epsilon = \frac{1}{f} \, . \label{eq:ekpyrosis}
\end{equation}
Slow contraction requires $\epsilon>3$, which corresponds to $f<1/3$ and in consequence to $V<0$. Slow contraction takes place from $t_i = -\infty$, since $V<0$ for all $\phi$. Note that with the power-law evolution one finds $\epsilon_N = \epsilon_{NN} = 0$, which automatically satisfies fast-roll conditions. Also worth mentioning is the fact that $f\gg1$ and $V>0$ is the so-called power-law inflation model.

\item[b)] Intermediate contraction: The model of Intermediate inflation \cite{Barrow:1993zq,Barrow:2006dh,delCampo:2009ma,Oikonomou:2017isf} is another example of exact analytical inflationary solution. This can be turned into a contracting scenario. In parts of the field space it is  responsible for slow contraction with an exact solution of the following form
\begin{equation}
a \propto e^{A(-t)^f} \, , \qquad V = \frac{8A^2}{(\beta+4)^2} \left(\frac{\phi}{\sqrt{2A\beta}}\right)^{-\beta} \left(6-\frac{\beta^2}{\phi^2} \right) \, , \qquad \epsilon = \frac{1}{f\, N} \label{eq:intermediateV}
\end{equation}
where $\beta = 4(1/f -1)$, $f \ll 1$ and $0 < A f \ll 1$. In this case $t_i$, as well as maximal number of e-folds of slow contraction are finite. The slow contraction starts when the potential becomes negative, which happens for $\phi < \beta/\sqrt{6}$ and $N < N_{\max} = \frac{1}{3f}$, where $N_{\max}$ is the maximal number of e-folds generated during the slow contraction phase of the Intermediate solution. Since $N_{\max} \geq 60$ one requires $f < 1/180$ in order to generate at least 60 e-folds of slow contraction. Since $fN = \epsilon^{-1}$ one can safely assume that throughout most of the slow contraction one finds $fN \ll 1$. This model in an excellent example of the fast-roll approximation, since $|\epsilon_N/\epsilon| = 1/N \ll 1$ and $|\epsilon_{NN}/\epsilon| = 2/N^2 \ll \epsilon$. 

In the $N\gg 1/f$ limit one finds $\epsilon \ll 1$, which corresponds to accelerated contraction of the Universe. For certain finely tuned initial conditions one can obtain a solution, in which $\phi$ stays for ever in the $V>0$ accelerated contraction regime. Nevertheless, $\phi$ naturally tends to roll towards negative potential for $f\ll 1$. Let us emphasize that during the phase of accelerated contraction, if it exists up until the bounce, the anisotropic stress may easily dominate the Universe and destroy the bounce. Therefore, the existence of the slow contracting phase is mandatory for obtaining a homogeneous contracting Universe.
\end{itemize}

\begin{figure}[H]
\centering
\includegraphics[width=0.31\textwidth]{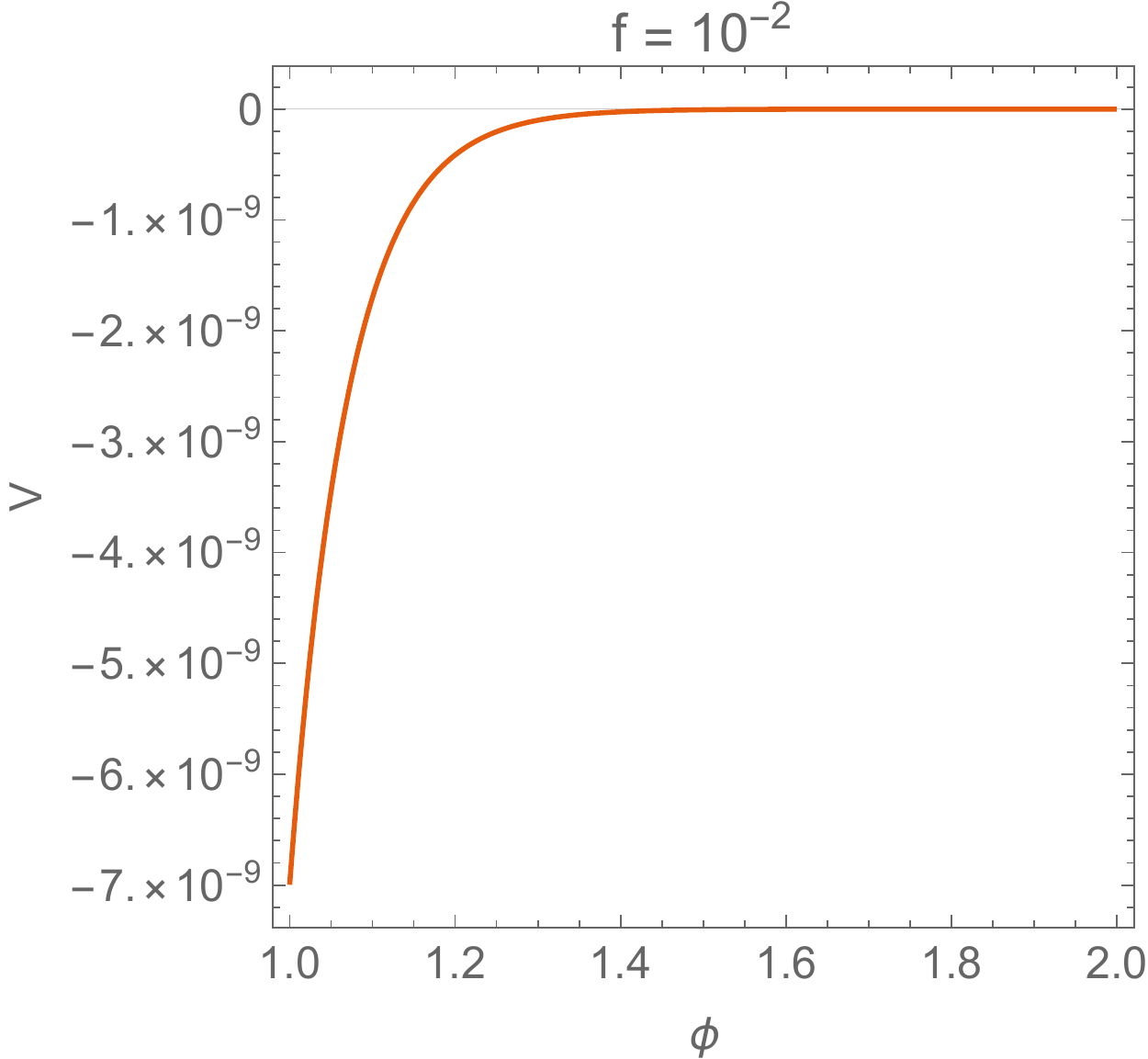} 
\hspace{0.25cm}
\includegraphics[width=0.31\textwidth]{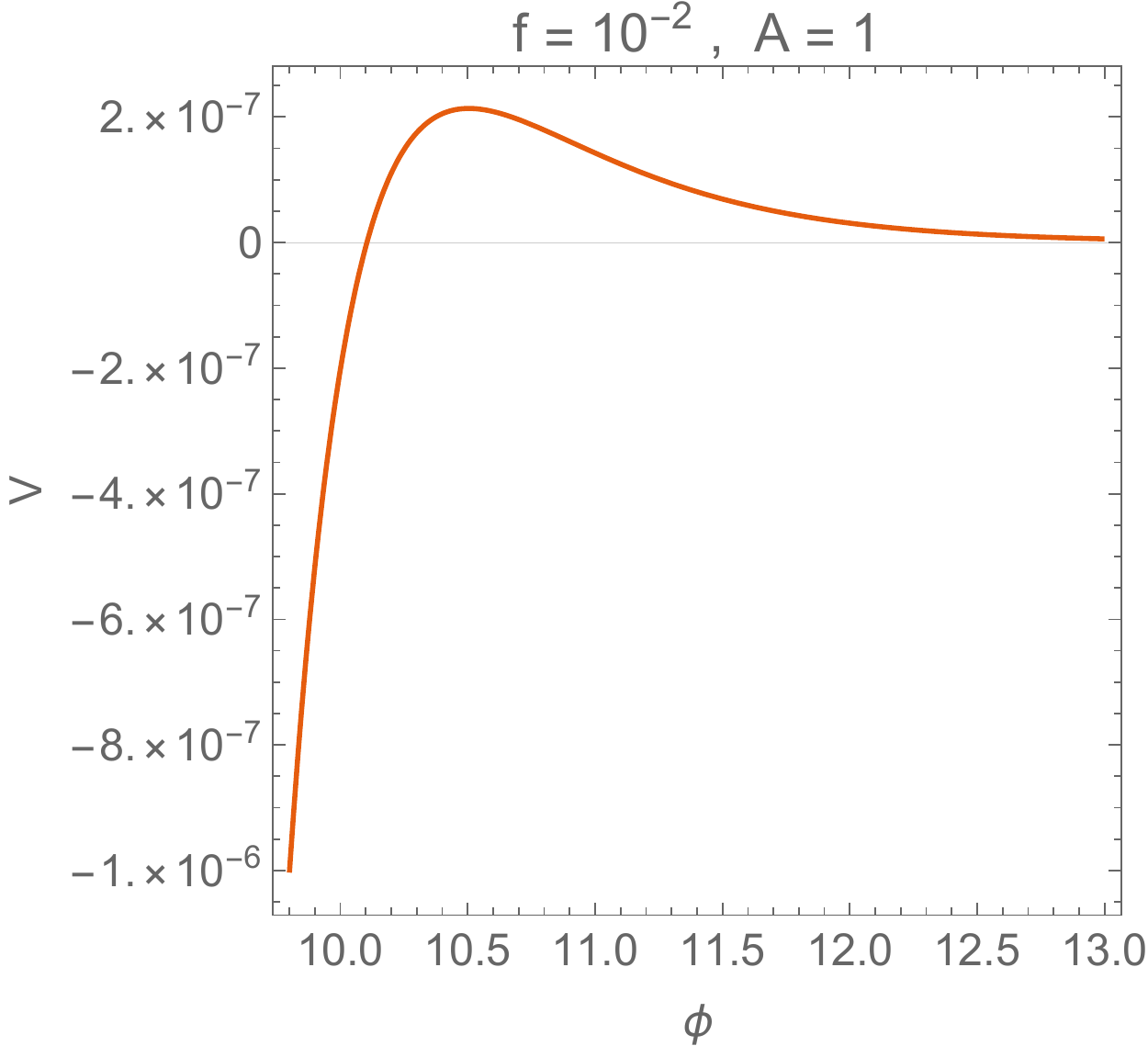}
\hspace{0.25cm}
\includegraphics[width=0.31\textwidth]{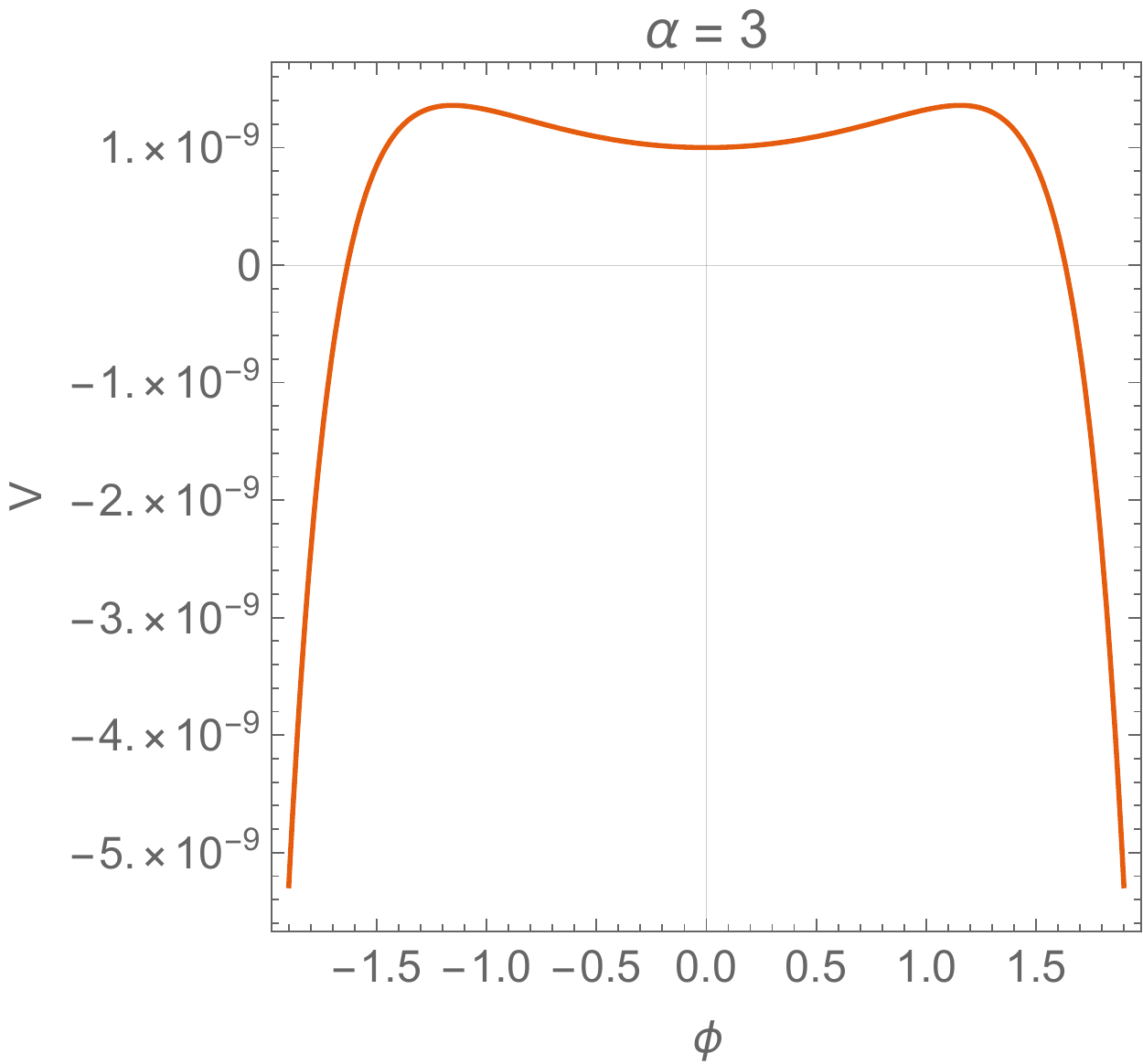} 
\caption{Left, middle and right panels show potentials for (\ref{eq:ekpyrosis},\ref{eq:intermediateV}) and \eqref{eq:expepsilon} models respectively. In ekpyrosis (left panel) $V<0$ for all $\phi$. For Intermediate and beyond fast-roll models (middle and right panel respectively) one obtains slow contraction in the $V<0$ regime and accelerated contraction in most of the $V>0$ regime. Note that accelerated contraction does not solve the problems of growing instabilities, since it leads to $-1<w<-1/3$. The energy density of the Universe redshifts too slow, which enables the domination of the $\rho \propto a^{-6}$ anisotropy term. Thus, a sufficiently long period of slow contraction is always required.} 
\label{fig:potentials}
\end{figure}

Slow contraction may also go beyond the fast-roll approximation. A good example of a model, which satisfies \eqref{eq:slowcontrgen}, but does not satisfy \eqref{eq:FRapprox} is 
\begin{equation}
\epsilon = 3 e^{-\alpha(N-N_i)} \, , \quad V(\phi) = V_0\,  e^{\frac{\alpha}{4}\phi^2}\left( 1 - \frac{\alpha^2\phi^2}{24} \right) \, , \quad \phi = \pm\frac{2\sqrt{6}}{\alpha}e^{-\frac{1}{2}\alpha(N-N_i)} \, , \label{eq:expepsilon}
\end{equation}
where $\alpha = const >0$. Note that \eqref{eq:expepsilon} is an exact solution of Friedmann equations, though we cannot write $a(t)$ explicitly in closed form. The slow contraction starts at $N = N_i$, when $\epsilon = 3$. The \eqref{eq:expepsilon} model satisfies general slow contraction conditions (i.e. Eqs. \eqref{eq:slowcontrcondition} and \eqref{eq:slowcontrgen}), but does not satisfy \eqref{eq:FRapprox}, since $|\epsilon_N/\epsilon| = \alpha$ and $\alpha$ does not have to be small. Potentials for models (\ref{eq:ekpyrosis}-\ref{eq:expepsilon}) are plotted in Fig. \ref{fig:potentials}.

\section{Vacuum perturbations and Green's functions} \label{sec:pertun}

Any given scalar field model of the contracting Universe determines not only its background evolution, but also the nature of primordial inhomogeneities of the scalar field and space-time. These primordial inhomogeneities are the seeds of the large scale structure of the Universe and a source of primordial gravitational waves. Thus, on top of the background evolution of $\phi(t)$ and $g_{\mu\nu} = diag(-1,a^2,a^2,a^2)$, one can consider \textit{quantum vacuum fluctuations} of these quantities of the form of $\delta\phi(\vec{x},t)$ and $\delta g_{\mu\nu}(\vec{x},t)$. Gauge invariant combinations of these perturbations can then be compared to observables which is explained in detail, for example in \cite{Langlois:2010xc}.

From measurements of CMB anisotropies \cite{Akrami:2018odb} we infer the power spectrum of curvature perturbations $\zeta$ and impose an upper bound on the tensor-to-scalar ratio $r$, which is the ratio between power spectra of primordial tensor perturbations $h$ and $\zeta$. Fourier modes of the first order scalar and tensor perturbations, denoted as $\zeta_k=v_S/z$ and $h_k = v_T/a$ respectively, satisfy Mukhanov-Sasaki equations:
\begin{equation}
v_i'' + \left( k^2-\frac{d_i''}{d_i}\right)v_i = 0 \, , \label{eq:MSeq}
\end{equation}
where $i=T,S$, $d_T=a$, $d_S = z$, $z = a\sqrt{2\epsilon}$ and $'$ denotes derivative with respect to conformal time given by $\tau = \int dt/a(t)$. For a given background one can solve \eqref{eq:MSeq}, which gives
\begin{equation}
\mathcal{P}_\zeta = \frac{k^3}{2\pi^2} \left| \frac{v_S}{z}  \right|^2\, , \qquad \mathcal{P}_h = 8 \frac{k^3}{2\pi^2} \left| \frac{v_T}{a}  \right|^2 \, . \label{eq:powerspectra}
\end{equation}
For effective model selection, a convenient parameterization is introduced:
\begin{equation}
r = \frac{\mathcal{P}_h}{\mathcal{P}_\zeta} \, , \qquad \mathcal{P}_\zeta \simeq A_\star \left( \frac{k}{k_\star} \right)^{n_s-1}
\end{equation}
where $A_\star$, is the amplitude of curvature perturbations, and $r$ is the tensor to scalar ratio evaluated at the so called pivot scale $k_*$ that somewhat varies between the different experiments. 
Scale dependence of power spectra is specified by scalar and tensor spectral indexes defined by
\begin{equation}
n_S = 1 + \frac{d\log \mathcal{P}_\zeta}{d\log k} \, , \qquad n_T = \frac{d\log \mathcal{P}_h}{d\log k}\, .
\end{equation}
The consistency between a given model and the data \cite{Akrami:2018odb} is determined by the value of $3$ parameters: $r$, $n_S$ and $A_\star$, all taken at the pivot scale $k_\star$. 

The superhorizon freeze-out of scalar and tensor perturbations, defined by $\zeta \to const.$ and $h \to const.$ for $-k\tau \ll 1$, is a fundamental feature of a successful theory of the primordial inhomogeneities. The lack of freeze-out may lead to the uncontrolled growth of perturbations, which can quickly reach the non-linear regime. While the superhorizon growth of scalar perturbations may be useful for producing primordial black holes \cite{Passaglia:2018ixg}, even in such a case the lack of freeze-out is temporary and it requires massive fine-tuning. Let us discuss the general conditions for the freeze-out of scalar and tensor modes and show how a realistic slow contraction secures the freeze-out. 

For tensors, the superhorizon solution to \eqref{eq:MSeq} is
\begin{equation}
h_k = c_1(k)+ c_2(k) \int \frac{dt}{a^3} = c_1(k)+ c_2(k) \int \frac{dN}{a^3H} \, , \label{eq:hsuper}
\end{equation}
where $c_1$ and $c_2$ are constants. Hence, a freeze-out requires the $c_2$ term to decrease during the evolution of the Universe. This is satisfied for any $a^{-3}$ decreasing faster than $t^{-1}$. Thus, in order to secure freeze-out one requires
\begin{equation}
\frac{d}{dt} \left| \frac{t}{a^3} \right| < 0 \quad \Rightarrow \quad |H| < \frac{1}{3|t|} \quad \Rightarrow \quad a < a_f \left( \frac{t}{t_f} \right)^{1/3} \, . \label{eq:htfreezeout}
\end{equation}
Condition \eqref{eq:htfreezeout} is automatically satisfied for any $\epsilon > 3$, which is equivalent to the general condition for the slow contraction \eqref{eq:slowcontrcondition}. Analogously, the superhorizon solution for $\zeta_k$ reads
\begin{equation}
\zeta_k = \tilde{c}_1(k)+ \tilde{c}_2(k) \int \frac{dt}{a^3 \epsilon} = \tilde{c}_1(k)+ \tilde{c}_2(k) \int \frac{dN}{a^3H \epsilon} \, . \label{eq:zetasuper}
\end{equation}
The freeze-out of scalar perturbations translates to the following condition
\begin{equation}
\frac{d}{dt} \left| \frac{t}{a^3\epsilon} \right| < 0 \quad \Rightarrow \quad \dot{\epsilon} > -3\left(H-\frac{1}{3t}\right)\epsilon \quad \Rightarrow \quad \epsilon > \epsilon_f \left( \frac{a}{a_f} \right)^{-3} \frac{t}{t_f}  \, . \label{eq:zetatfreezeout}
\end{equation}
and in terms of efolds:
\begin{equation}
\frac{d}{dN}\left( \frac{N}{a^3H \epsilon} \right) > 0 \qquad \Rightarrow \qquad \epsilon > 3 - \frac{1}{N} + \frac{\epsilon_N}{\epsilon} \, . \label{eq:scalarfreezecond}
\end{equation}
Since $-3(H-\frac{1}{3t})\epsilon <0$ one always satisfies \eqref{eq:zetatfreezeout} for any $\dot{\epsilon}>0$ (i.e. $\epsilon_N<0$). Generally speaking one should expect $\epsilon_N < 0$ during any realistic form of slow contraction, since it starts at some $t=t_i$, when $\epsilon = 3$ and $V(\phi) = 0$. The scalar field that is responsible for the slow contraction rolls downhill\footnote{In the framework of modified gravity the uphill evolution is also possible \cite{Artymowski:2018ewb}. Nevertheless, in this work we restrict ourselves to general relativity.}, into the $V < 0$ regime, which increases the value of $\epsilon$. Thus, any realistic slowly contracting model should satisfy \eqref{eq:zetatfreezeout}. On top of that, Eq \eqref{eq:zetatfreezeout} is always satisfied within the fast-roll approximation, since $\left|\epsilon_N/\epsilon \right| \ll 1$ is equivalent to $|\dot{\epsilon}/\epsilon|\ll |H|$. This gives \eqref{eq:zetatfreezeout} of the form of $H-\frac{1}{3t} > 0$, which is fully equivalent to \eqref{eq:htfreezeout}. 

To fully infer the predicted spectrum, we need to specify $a(t)$. Nevertheless, we can comment on differences between scalar and tensor perturbations within the slow contraction limit. For $\frac{z''}{z} \simeq \frac{a''}{a}$ one expects the same solutions for $v_S$ and $v_T$. Note that
\begin{equation}
\frac{a''}{a} = \mathcal{H}^2(2-\epsilon) \, , \qquad \frac{z''}{z} = \mathcal{H}^2 \left( (2-\epsilon)+ \frac{\epsilon_{N N}}{2\epsilon}-\frac{1}{4} \left(\frac{\epsilon_{N }}{\epsilon}\right)^2+ \frac{3\epsilon_N}{2 \epsilon} - \frac{1}{2} \epsilon_N \right)  \, . \label{eq:appzpp}
\end{equation}
Since $\epsilon \gg 1$, the leading terms in $(\frac{z''}{z}-\frac{a''}{a})\frac{a}{a''}$ are of order of $\epsilon_N/\epsilon$, $\epsilon_N^2/\epsilon^3$ and $\epsilon_{NN}/\epsilon^2$, all of which are much smaller than unity within the fast-roll approximation. Thus, for every model, which satisfies \eqref{eq:FRapprox} one finds $v_T \simeq v_S$. Note that this approximation does not work for models beyond fast-roll, e.g. for \eqref{eq:expepsilon}. 
The fact that $v_T\simeq v_S$ in the fast-roll approximation is crucial as it fixes not just the vacuum spectrum, but also the Green's functions necessary for the calculation of the sourced spectrum, limiting possible outcomes. We shall use this approximation in our analysis. We defer the analysis of $v_S\neq v_T$ scenarios to future work.

The vacuum power spectrum of both scalar and tensor perturbations during the slow contraction tends to be strongly blue-tilted and therefore inconsistent with the data \cite{Akrami:2018odb}. Eq. \eqref{eq:MSeq} can give scale invariant solutions only in 2 cases: For $a \propto \tau^2$ (the matter bounce) and for $a \propto \tau^{-1}$ (the cosmic inflation), or a linear combination of the two \cite{Wands:1998yp}, none of which satisfies slow contraction conditions. Hence, for viable slow-contracting models another scalar field is introduced to obtain the observed nearly scale invariant scalar spectrum via the entropic/curvaton mechanism \cite{Lehners:2010fy,BATTEFELD20151}. The tensor spectrum remains blue -unobservable on CMB scales, and detectable by Laser Interferometers \cite{Boyle:2007zx}.

\subsection{Sourced fluctuations}\label{subs:sf}

An alternative route for obtaining scale invariant spectrum for a contracting Universe was suggested in \cite{r1,r3}. A source term in \eqref{eq:vsourced} induces additional spectra for both the scalar and tensor modes. Hence, it is possible to modify $n_S$ and $n_T$ in such a way that power spectrum becomes nearly scale invariant.  The example considered in \cite{r1,r3} was a gauge field that is dynamically coupled to the scalar bouncer field. Introducing such gauge fields can source second order inhomogeneities that dominate over the blue tilted first order perturbations. Following \cite{Barnaby:2012xt,r1,r3} equation \eqref{eq:MSeq} is modified by a source term: 
\begin{equation}
v_i'' + \left( k^2-\frac{d_i''}{d_i}\right)v_i =  J_i(\tau,\Vec{k} ) \, , \label{eq:vsourced}
\end{equation}
where $J_i(\tau',\Vec{k} )$ is the corresponding source term that depends on the particular model. Since Eq. \eqref{eq:vsourced} is linear in $v_i$ and creation/annihilation operators of sourced and unsourced fluctuations are uncorrelated \cite{Barnaby:2012xt}, its solutions and their power spectra $\mathcal{P}$ should be linear combinations of vacuum and sourced fluctuations 
\begin{equation}
v_i =  v_{i,v} + v_{i,s}  \qquad \Rightarrow \qquad \mathcal{P}^{tot} = \mathcal{P}^v + \mathcal{P}^s \, 
\end{equation}
where $ v_{i,v} $ is the solution to homogeneous equation given by Eq. \eqref{eq:MSeq} and  
\begin{equation}
v_{i,s} = \int d\tau' G^i_{k}(\tau,\tau') J_i (\tau',\Vec{k}) \, ,
\label{eq:vi} 
\end{equation}
where $G^i_k(\tau,\tau')$ is the Green's function. $G^S_k$ and $G^T_k$ are scalar and tensor Green's functions defined by
\begin{equation}
G^{ \prime \prime i} + \left(k^2 - \frac{d_i''}{d_i}\right)G^{i} = \delta(\tau-\tau') \, .
\label{eq:greenf}
\end{equation}
The sourced two point function is then given by the formal expression:
\be
\langle \hat X_k(\tau) \hat X_{k'}(\tau) \rangle= \int d\tau' \frac{G_{k}^i(\tau,\tau')}{d_i(\tau)} \int d\tau'' \frac{G_{k'}^i(\tau,\tau'')}{d_i(\tau)} \langle \hat J_i (\tau',\vec{k}) \hat J_i (\tau'',\vec{k'}) \, \rangle
\label{eq:generalsf}
\ee
where $X_k=h_k,\zeta_k$  where $d_i =a,z$ for the tensor and scalar perturbations respectively. Tensor to scalar ratio for general sourced perturbations is given by
\begin{equation}
r = \left(\frac{a(\tau)}{z(\tau)}\right)^2 \frac{ \langle v_T(k,\tau) v_T(k',\tau)  \rangle}{\langle v_S(k,\tau) v_S(k',\tau)  \rangle} = \frac{\int d\tau' \frac{G_{k}^T(\tau,\tau')}{a(\tau)} \int d\tau'' \frac{G_{k}^T(\tau,\tau'')}{a(\tau)} \langle J_T (\tau',\vec{k}) J_T (\tau'',\vec{k'}) \, \rangle }{\int d\tau' \frac{ G_k^S(\tau,\tau')}{z(\tau)}  \int d\tau''  \frac{G_k^S(\tau,\tau'')}{z(\tau)} \langle J_S (\tau',\vec{k}) J_S (\tau'',\vec{k'}) \, \rangle} \, . \label{eq:r}
\end{equation}
Let us find a general functional form of Green's functions for slow contracting Universe at superhorizon scales.
Green's functions, which are solutions to Eq. (\ref{eq:greenf}) are of the form 
\begin{equation}
G(\tau,\tau') = \frac{1}{W(y_1,y_2)}\Theta (\tau-\tau') \left(y_1(\tau) y_2(\tau') - y_1(\tau') y_2(\tau)\right)
\label{eq:Green'sg}
\end{equation}
where $y_1$,$y_2$ are solutions to the homogeneous equation (i.e. solutions to Eq.\eqref{eq:MSeq}) and $W$ is the Wronskian.
For tensor perturbations, in the superhorizon limit one finds
\begin{equation}
y_1 = c_1(k) a  \qquad y_2 = c_2(k) a \int \frac{d\tau}{a^2} \, .
\end{equation}
Quantization of perturbations necessitate $|W|=1$ \cite{Baumann:2009ds}. Hence, the Green's function in the superhorizon limit is given by
\begin{equation}
G_k^T(\tau,\tau') = \Theta (\tau-\tau') c_1(k)c_2(k) a(\tau) a(\tau')  \int^{\tau'}_{\tau} \frac{d u}{a(u)^2} \, . \label{Green'st}
\end{equation}
From dimensional analysis one can see that the full solutions $y_1$ and $y_2$ should be of the form $\frac{1}{\sqrt{k}} z(x)$ for some function $z(x)$, where $x=-k \tau$. Thus, the functional form of Green's functions at super horizon scales and $\abs{\tau}<\abs{\tau'}$ is
\begin{equation}
G_k^T(\tau,\tau') \propto a(\tau) a(\tau') \frac{g(k \tau'))}{k} \, , \quad \text{where} \quad g(x) \propto \int \frac{dx}{a(x)^2} \, .\label{Green's}
\end{equation}
The Green's function for scalars can be expanded in a similar fashion for super horizon scales:
\begin{equation}
G_k^S(\tau,\tau') \propto \Theta (\tau-\tau') \tilde{c}_1(k)\tilde{c}_2(k) \sqrt{\epsilon(\tau)} \sqrt{ \epsilon(\tau')} a(\tau) a(\tau')  \int^{\tau'}_{\tau} \frac{d \eta}{a(\eta)^2\epsilon(\eta)}
\label{Green's1}.
\end{equation}
Functional form of scalar Green's function at super horizon scales  and $\abs{\tau}<\abs{\tau'}$ can be expressed similar to Green's function for tensors \eqref{Green's}.
\begin{equation}
G_k^S(\tau,\tau') \propto \sqrt{\epsilon(\tau)} \sqrt{ \epsilon(\tau')}a(\tau) a(\tau') \frac{g(k \tau')}{k} \, , \quad \text{where} \quad \tilde{g}(x) \propto \int \frac{dx}{\epsilon(x) a(x)^2} \, .
\label{Green'ss}
\end{equation}
Summarizing, for super horizon scales, up to first order in $x$: 
\begin{eqnarray}
G_k^T(\tau,\tau') &\propto& a(\tau) a(\tau') g'(\tau) (\tau'-\tau) = \frac{a(\tau)}{a(\tau')}(\tau'-\tau) \, ,  \\
G_k^S(\tau,\tau')  &\propto &  \sqrt{\frac{\epsilon(\tau)}{\epsilon(\tau')}}a(\tau) a(\tau') g'(\tau) =\frac{z(\tau)}{z(\tau')}(\tau'-\tau)\, .
\label{eq:Green'sf}
\end{eqnarray}
i.e. the Green's functions of both the scalar and tensor fluctuations are independent of $k$.

\subsection{Green's functions for fast-roll scenarios} \label{sec:Green'sfi}

We have established that Green's functions for superhorizon scales are independent of the scale k. It is however not possible to determine the exact form of Green's functions without knowledge of the background dynamics. Nevertheless, one can compare the tensor and scalar Green's functions within the fast-roll approximation. As argued in Eq. \eqref{eq:appzpp} $\frac{a''}{a} \simeq \frac{z''}{z}$ within  fast-roll. Since Green's functions are solutions to Eq.\eqref{eq:Green'sf}, which are the same within fast-roll and have the same vacuum boundary conditions 
\begin{equation}
    G_k^T(\tau,\tau')_{fast-roll} \simeq G_k^S(\tau,\tau')_{fast-roll} = G_k(\tau,\tau')
    \label{eq:gffastroll}
\end{equation}
To further explore the form of perturbations and Green's functions in a fast-roll scenario We define parameters $f_i$ as 
\begin{equation}
f_0 = a \, , \qquad f_1 \equiv \frac{dN}{d\log \tau} = \tau\, \mathcal{H} \, , \qquad f_{2} \equiv \frac{d \log f_1}{d\log \tau}\, , \qquad f_{i+1}\equiv \frac{d \log f_i}{d\log \tau}\, , \quad \cdots . \label{eq:fi}
\end{equation}
Since we assume $\tau < 0$ in the contracting Universe, one finds $f_1>0$ throughout the whole period of contraction. Furthermore, Eq. \eqref{eq:fi} can be expressed using the number of e-folds as a time variable, which gives $f_{2} = f_{1,N}$. Inequality $\epsilon > 3$ takes the form
\begin{equation}
\epsilon -1 =-\frac{\mathcal{H}'(\tau)}{\mathcal{H}^2} = \frac{1-f_2}{f_1} > 2 \, . \label{epsilon}
\end{equation}
Combining Eqs.
\eqref{epsilon} and \eqref{eq:appzpp} and truncating the expansion of $\frac{z''}{z}$ at the largest term $-\frac{\epsilon_N}{2}$
\begin{equation}
\frac{a''}{a} = \frac{f_1}{\tau^2} (-1+f_1+f_2) 
\end{equation}
and
\begin{equation}
\frac{z''}{z} = \frac{1}{\tau^2} (f_1(-1+f_1+f_2)+\frac{f_2}{2}(1-f_3-f_2)).
\end{equation}
For fast-roll $f_i\ll1$, thus
\begin{equation}
\frac{a''}{a} \simeq -\frac{f_1}{\tau^2}  \,\,,\,\,  \qquad   \frac{z''}{z} \simeq \frac{f_2/2-f_1}{\tau^2}.
\end{equation}
For fast-roll $f_i$s are slowly varying with $\tau$, and can be assumed to be constants for the purpose of solving Eq.\eqref{eq:MSeq}, whose solutions are given by
\begin{equation}
v_S  \simeq \frac{\sqrt{\pi (-\tau)}}{2} H_{\frac{1}{2} + f_1}^{(1)}(k\tau) \, , \qquad v_T  \simeq \frac{\sqrt{\pi (-\tau)}}{2} H_{\frac{1}{2} +f_1-f_2/2}^{(1)}(k\tau)
\label{eq:vacfs}
\end{equation}
where  $H_{m}^{(1)}$ is Hankel function of first kind. From Eq.\eqref{eq:Green'sg}, tensor and scalar Green's functions are given 
\begin{equation}
\begin{split}
G^T_k(\tau,\tau') &= i \Theta(\tau-\tau')  \frac{\pi}{4} \sqrt{\tau \tau'}
\left[H^{(1)}_{\frac{1}{2}+f_1}(-k\tau) H^{(2)}_{\frac{1}{2}+f_1}(-k\tau')-H^{(1)}_{\frac{1}{2}+f_1}(-k\tau')H^{(2)}_{\frac{1}{2}+f_1}(-k\tau)\right],\cr
\end{split}
\label{eq:geensfsten}
\end{equation}
\begin{equation}
\begin{split}
G^S_k(\tau,\tau') &= i \Theta(\tau-\tau')  \frac{\pi}{4} \sqrt{\tau \tau'} \times\\&\,
\left[H^{(1)}_{\frac{1}{2}+f_1-\frac{f_2}{2}}(-k\tau) H^{(2)}_{\frac{1}{2}+f_1-\frac{f_2}{2}}(-k\tau')-H^{(1)}_{\frac{1}{2}+f_1-\frac{f_2}{2}}(-k\tau')H^{(2)}_{\frac{1}{2}+f_1-\frac{f_2}{2}}(-k\tau)\right].
\end{split}
\label{eq:geensfsten2}
\end{equation}
Expanding this Green's functions outside the horizon ($-k\tau\ll1$) and for $f_1,f_2 \ll 1$
\begin{equation}
     G_k^T(\tau,\tau') \simeq G_k^S(\tau,\tau') \simeq  \Theta(\tau-\tau')   \frac{\sin(-k\tau')}{k} .
     \label{eq:greenfss}
\end{equation}
Let us consider the $-k\tau'\ll1$ limit, as it provides the biggest contribution to the sourced perturbations. From \eqref{eq:greenfss} one finds $G_k(\tau,\tau') \simeq -\Theta(\tau-\tau') \tau' $, which is independent of $k$. This result is fully consistent with the analysis from Sec. \ref{subs:sf}. Small deviations from this equality will yield that $G_T \geq G_S$ , see appendix \ref{app:Green'scomp}. Expanding the Green's functions \eqref{eq:geensfsten} and  \eqref{eq:geensfsten2} for $-k\tau\ll1$, we observe $G_S,G_T>0$, combining these observations with \eqref{eq:r} tells us that r is smallest when $G_T=G_S$. 


\section{U(1) Gauge field coupled to a scalar field } \label{sec:gauge}

Let us consider the following action for our contracting model:
\begin{equation}
S = \int d^4x \sqrt{-g} \left[R-\frac{1}{2}\left(\partial \phi\right)^2-V(\phi)-I^2(\phi) \left( \frac{1}{4}F^{\mu\nu}F_{\mu \nu} -\frac{\gamma}{4}\tilde{F}^{\mu \nu}F_{\mu\nu}\right) \right] \label{eq:vecaction}
\end{equation}
where $\phi$ is the bouncer field, $A_{\mu}$ is the U(1) gauge potential,  $F_{\mu\nu} = \partial_\mu A_\nu - \partial_\nu A_\mu$, $\tilde{F}^{\mu\nu} = \frac{1}{2}\epsilon^{\mu\nu\rho\sigma}F_{\rho\sigma}$,  $\gamma>0$ is a coupling constant that can be a VEV of some other field,  and $V(\phi)$ is some potential that generates slow contraction. We shall work in the Coulomb gauge, namely $\partial^i A_i = A_0=0$. The gauge field can be decomposed into the Fourier modes by
\begin{equation}
\vec{A}(\tau,{\vec x}) = \sum_{\lambda=\pm} \int \frac{d^3k}{(2\pi)^{3/2}} \left[ \vec{\epsilon}_\lambda({\vec k}) \hat{a}_{\lambda}({\vec k}) A_\lambda(\tau,{\vec k}) e^{i {\vec k}\cdot {\vec x}} + \mathrm{h.c.}   \right] \, ,
\end{equation}
where $\vec{\epsilon}_\lambda$ are polarization vectors, which satisfy $\vec{k}\cdot \vec{\epsilon}_{\pm} \left( \vec{k} \right) = 0$, 
$\vec{k} \times \vec{\epsilon}_{\pm} \left( \vec{k} \right) = \mp i k \vec{\epsilon}_{\pm} \left( \vec{k} \right)$,
$\vec{\epsilon}_\pm \left( -\vec{k} \right) = \vec{\epsilon}_\pm \left( \vec{k} \right)^\star$ and  $\vec{\epsilon}_\lambda \left( \vec{k} \right)^\star 
\cdot \vec{\epsilon}_{\lambda'} \left( \vec{k} \right) = \delta_{\lambda \lambda'}$. 
The annihilation and creation operators obey $  \comm{\hat{a}_{\lambda}}{\hat{a}^{\dagger}_{\lambda}} = \delta_{\lambda \lambda'} \delta^3(\vec{k}-\vec{k'})$. The essential feature of the model is that annihilation and creation operators of the gauge field commute with the operators of the tensor and scalar fluctuations. This leads to the lack of mixing terms between sourced and vacuum perturbations.
Following Ref. \cite{Caprini:2014mja,r1,r3} we define $\tilde{A}=IA$, where $\tilde{A}$ satisfy the equation of motion of the form 
\begin{equation}
\tilde{A''}_{\lambda}+\left(k^2+2\lambda \gamma k \frac{I'}{I} -  \frac{I''}{I}\right) \tilde{A}_{\lambda} = 0 \, , \label{guage}
\end{equation}
where $\lambda=\pm 1$ denote polarization of the field and $I' = \frac{dI}{d\tau}$. The crucial feature of \eqref{guage} is the fact that it is completely background independent. Thus, $A_\lambda$ and $\tilde{A}_\lambda$ depend only on $I(\tau)$ and \emph{not} on a particular form of $a(\tau)$. It allows us to investigate the evolution of the gauge field without narrowing down the background evolution to a particular model. The form of $I(\tau)$ will be essential for the scale-invariance of primordial inhomogeneities. 

Gauge field can be a source of second order inhomogeneities, which in the case of sourced perturbations may be the leading order. The particular form of the source term considered in this work is well described in the literature in the context of the power-law contraction \cite{r1,r3}. In particular, Ref. \cite{r1} contains detailed analysis of perturbations of both scalar and vector fields up to the second order. 
Since the energy momentum tensor is quadratic in gauge fields one finds
\begin{equation}
J_{\lambda}(\tau',\Vec{k} ) = \int \frac{ d^3p}{(2\pi)^\frac{3}{2}} \hat{O}_{\lambda ij} (\tau,\Vec{k},\Vec{p}) \tilde{A}_i(\tau,\Vec{p}) \tilde{A}_j (\tau, \vec{k}-\vec{p}).
\end{equation}
for some operator $\hat{O}$ dependent on the source and helicity $\lambda$. 
In \cite{r1,r3} it was shown that only one  polarization contributes significantly to the power spectrum, due to its exponential enhancement. Such enhancement implies that the magnitude of sourced perturbations can be several orders of magnitude larger than that of vacuum perturbations. Thus, the observed power spectrum will simply be the sourced spectrum.
$J_{\lambda}^T (\tau,\Vec{k})$ is obtained by taking transverse and traceless spatial part of the energy momentum tensor and projecting it along the $\lambda$ polarization. For the action given in Eq. \eqref{eq:vecaction} \cite{r3}
\begin{equation}
\begin{split}
J_{\lambda}^T (\tau,\Vec{k}) &\simeq -\frac{1}{2 a}\int \frac{d^3p}{2\pi^3} \sum_{\lambda'} \epsilon_i^{\lambda*}(\vec{k}) \epsilon_j^{\lambda*}(\vec{k})
\epsilon_i^{\lambda'}(\vec{p})\epsilon_j^{\lambda'}(\vec{p}-\vec{k}) \frac{\left(\tilde{A}'_{\lambda} I -\tilde{A}_{\lambda} I'\right)^2}{I^2}\\
&\;\;\;\;\;\;\;\; \times \left[ \hat{a}_{\lambda'}(\vec{p})+\hat{a}^{\dagger}_{\lambda'}(-\vec{p})\right]\left[ \hat{a}_{\lambda'}(\vec{k}-\vec{p})+\hat{a}^{\dagger}_{\lambda'}(-\vec{k}+\vec{p})\right],
\end{split}
\label{eq:tensource}
\end{equation}
while the source term for scalar perturbations is given by
\begin{equation}
\begin{split}
J_{\lambda}^S (\tau,\Vec{k})\simeq & \frac{1}{2 a}\int \frac{d^3p}{2\pi^3} 
\epsilon_i^{\lambda}(\vec{p})\epsilon_j^{\lambda}(\vec{p}-\vec{k}) \frac{\left(\tilde{A}'_{\lambda} I -\tilde{A}_{\lambda} I'\right)^2}{I^2}\\
&\;\;\;\;\;\;\;\; \times C(\tau) \sqrt{2\epsilon(\tau)} \left[ \hat{a}_{\lambda}(\vec{p})+\hat{a}^{\dagger}_{\lambda}(-\vec{p})\right]\left[ \hat{a}_{\lambda}(\vec{k}-\vec{p})+\hat{a}^{\dagger}_{\lambda}(-\vec{k}+\vec{p})\right].
\end{split}
\label{eq:scsource}
\end{equation}
After extracting all time-dependence from \eqref{eq:tensource} and \eqref{eq:scsource} one obtains a momentum-dependent functions, which in the further parts of this paper will be called momentum integrals. The particular form of momentum integrals determines the $k$-dependence of power spectra and it strongly depends on the solution of the gauge field. In this model, $J_{\lambda}^T (\tau,\Vec{k})$ and $J_{\lambda}^S (\tau,\Vec{k})$ are the same up to a projection tensor and a time dependent function $C(\tau)$:
\begin{equation}
C(\tau)= \frac{\mathcal{H}}{\phi'}\left[\dfrac{dI^2}{d\phi}+ \frac{\phi'}{\mathcal{H}} I^2 \right] I^{-2}(\phi) =  1-\frac{I'}{I} \frac{1}{\epsilon \mathcal{H}} = 1 - \frac{d\log I}{d \log H} 
\label{eq:C}
\end{equation}
Let us note that the similarity of the source field is solely due to the fact that the electric field of the $U(1)$ gauge field is the dominant term for both the scalar and tensor perturbations. This does not have to be the case with other sources coupled to the bouncer field. Having source terms and Green's functions one can obtain power spectra of primordial inhomogeneities. 


\subsection{Nearly scale invariant sourced spectrum} \label{sec:scaleinv}

Our aim is to determine the form of $\tilde{A}$ that secures scale invariance of the spectrum without assuming a particular form of the scale factor. Correlation functions for  curvature and tensor sourced perturbations can be obtained by substituting Eqs.\eqref{eq:tensource} and \eqref{eq:scsource} in Eq.\eqref{eq:generalsf}
\begin{eqnarray}
&\langle \hat{h}_{\lambda,s} \hat{h}_{\lambda',s} \rangle = \int d\tau' \frac{G^T_k(\tau,\tau')}{a(\tau) a(\tau')} \int d\tau'' \frac{G^T_k(\tau,\tau'')}{a(\tau) a(\tau'')} \langle J^T_{\lambda}(\vec{k},\tau') J^T_{\lambda'}(\vec{k'},\tau'') \rangle ,
\label{eq:factorizationt} \\
&\langle \hat{\zeta}_{\lambda,s}\hat{\zeta}_{\lambda',s} \rangle  = \int d\tau' \frac{G^S_k(\tau,\tau')}{z(\tau) a(\tau')} \sqrt{2 \epsilon(\tau')} \,C(\tau') 
\int d\tau'' \frac{G^S_k(\tau,\tau'')}{z(\tau) a(\tau'')} \sqrt{2 \epsilon(\tau'')} \, C(\tau'')\langle J^S_{\lambda}(\vec{k},\tau') J^S_{\lambda'}(\vec{k'},\tau'') \rangle .\cr
\label{eq:factorizations} 
\end{eqnarray}
At the superhorizon limit gauge fields are of the form
\begin{equation}
    \Tilde{A_{\lambda}}(k,\tau) = D_1(k) I(x) + D_2(k) I(x) \int \frac{1}{I(x)^2} dx,
    \label{subhg}
\end{equation}
where $x=-k\tau$.
Substituting for $\Tilde{A}_{\lambda}$, the time independent $D_1(k)$ term drops out and we are left with 
\begin{equation}
\begin{split}
J_{\lambda}^S (\tau,\Vec{k})\simeq & \frac{1}{2 a I(x)^2}\int \frac{d^3p}{(2\pi)^3} 
\epsilon_i^{\lambda}(\vec{p})\epsilon_j^{\lambda}(\vec{p}-\vec{k}) \left(D_2(p)D_2(\abs{\Vec{k}-\vec{p}}) \right) k^2\\
&\;\;\;\;\;\;\;\; \times C(\tau) \sqrt{2\epsilon(\tau)} \left[ \hat{a}_{\lambda}(\vec{p})+\hat{a}^{\dagger}_{\lambda}(-\vec{p})\right]\left[ \hat{a}_{\lambda}(\vec{k}-\vec{p})+\hat{a}^{\dagger}_{\lambda}(-\vec{k}+\vec{p})\right] \, ,
\end{split}
\label{eq:scsourcegen}
\end{equation}
where the $k^2$ term inside the integral comes from changing the  differentiation with respect to $\tau$ to a differentiation with respect to $x$. As we prove in Appendix \ref{app:subh}, subhorizon contributions cannot give a scale invariant spectrum. Scale invariance will come from the superhorizon contributions. Combining Eq. \eqref{eq:factorizations} with the form of Green's functions for fast-roll from discussed in Sec. \ref{sec:Green'sfi}: 
\begin{equation}
\begin{split}
 \langle \hat{\zeta}_{\lambda,s}\hat{\zeta}_{\lambda',s} \rangle &\propto \left(\int_{x_{end}}^{1} dx' x'  \frac{ \sqrt{2\epsilon} C(x')}{a(x')} \frac{1}{ k^2 I(x')^2} \right)^2 k^{4} \times  \\
 &\int \frac{d^3p}{(2\pi)^3}\frac{1}{\sqrt{2p\abs{\vec{k}-\vec{p}}}} \int \frac{d^3p'}{(2\pi)^3} \delta(\vec{p}-\vec{p'}) \frac{1}{\sqrt{2p'\abs{\vec{k}-\vec{p'}}}},
 \end{split}
\end{equation}
after using Bunch-Davies vacuum boundary conditions $D_1(k)=D_2(k)=\frac{1}{\sqrt{2k}}$. Thus the scalar power spectrum up to some numerical factor is
\begin{equation}
    \mathcal{P}_S \propto \left(k^4 \int_{x_{end}}^{1} x' dx'  \frac{ \sqrt{2\epsilon} C(x')}{a(x')} \frac{1}{k^2 I(x)^2} \right)^2 .
\end{equation}
In order for the power spectrum to be scale invariant the integral over $x$ should contribute a factor of $k^{-8}$. For fast-roll, contributions from all other factors in the integral except I are practically constant. Scale invariance hence requires $I(x)$ to be of power law form. For $I(x)=x^{-n}$, rigorous calculations with the full equations for gauge field will show that the dominant contribution to the integral in super-horizon limit is coming from the interval $x_{end}<x<\frac{1}{\xi}$ \cite{r1,r3}, where $\xi=\lambda n \gamma$. The largest contribution to the time integral is then given by 
\begin{equation}
    \mathcal{I} = \int_{x_{end}}^{\frac{1}{\xi}} x' dx'  \frac{ \sqrt{2\epsilon} C(x')}{a(x')} \frac{1}{k^2 I^2} \propto  k^{2n} \tau_{end}^{2+2n}
\end{equation}
Hence in this case k dependence of the power spectrum is\footnote{For inflation given $n\geq -2$,
\begin{equation}
    \mathcal{I} = \int_{x_{end}}^{\frac{1}{\xi}} x^{' 3} dx'  \frac{ \sqrt{2\epsilon} C(\tau')}{a(\tau)} \frac{1}{I^2} = \frac{1}{(k\xi)^4} 
\end{equation}
and power spectrum is scale independent for any $n$.}
\begin{equation}
     \mathcal{P} \propto k^{8+4n}. 
\end{equation}
This is all in accord with known results. We have determined that scale invariance require powerlaw for of $I$. For power law form of $I$ the scale dependence of the momentum integrals are the same as $f^T$ and $f^S$ defined in \cite{r1,r3}. Since for power law  the integral over $\tau$ does not alter the $k$-dependence of perturbations \citep{r1,r3}, scale dependence is solely determined by the momentum integral. The difference between the momentum integrals of the tensor and scalar perturbations is only in the projection tensor, hence both of them have the same scale dependence. Consider for example the momentum integral in the scalar correlation function $\langle \hat \zeta \hat \zeta \rangle$. Let us determine the scale dependence of the gauge field that will provide a scale invariant spectrum. 
\begin{equation}
\begin{split}
&\bigg \langle \hat{\zeta}_{\lambda}(\hat{\lambda},\vec{k})\hat{\zeta}_{\lambda}(\hat{\lambda'},\vec{k'})  \bigg \rangle \propto k^{3+4m} \int \frac{d \cos(\theta) q^2dq}{(2\pi)^2} \abs{\epsilon_i\epsilon_i}^2  \left(q(\abs{\hat{k}-\vec{q}})\right)^{2n} \, .
\end{split}
\label{diagram}
\end{equation}
where $q\equiv \vec{p}/k$. Hence the power spectrum $P_T \propto P_S \propto k^{4m+6}$. Requiring scale invariance necessitate that $m=-\frac{3}{2}$. We want to emphasize that  $k^{-3/2}$ is the only form of the $k$-dependence of $\tilde{A}$, which allows spectra to be scale independent as long as the scale dependence is arising from momentum integral alone. The superhorizon solutions of $\tilde{A}$ with $k$-dependence of the form of $k^{-3/2}$ are well known in the literature \cite{Caprini:2014mja} and they require:
\begin{equation}
\frac{I''}{I} = \frac{2}{(-\tau)^2} \qquad \Rightarrow \qquad I = I_1 (-\tau)^2 + I_2 (-\tau)^{-1} \, , \label{eq:Iscaleinv}
\end{equation}
where $I_1$ and $I_2$ are constants. Thus, it is safe to conclude that the only $I(\tau)$ that can give a scale invariant power spectrum for sourced primordial inhomogeneities is $I \propto (-\tau)^{-n}$, where $n=1$ or $n=-2$.\footnote{Considering this form of $I(\tau)$ there is actually a duality in the action and equations of motion: $n\rightarrow -1-n$ and $\gamma\rightarrow -\gamma n/(1+n)$. This duality is only broken in the scalar perturbations in $C(\tau)$, \cite{r1,r3}.} 

Let us discuss the implications of this result. We have trodden two paths in the effort of producing a viable Sourced Bounce. First, we have allowed a general coupling $I(\tau)$ between the gauge field and the bouncer. Second, we consider general slow contraction, not necessarily ekpyrosis. Our analysis shows that the requirement of scale invariance limits the general coupling to be \eqref{eq:Iscaleinv}, which was already discussed in \cite{r1,r3} and is a very strong limitation. We shall see that for the $U(1)$ gauge field, fixing $I$ actually fixes $r\simeq 1/9$ provided that the fast-roll conditions are met.   

\section{Tensor to scalar ratio for sourced perturbations} \label{sec:r}

For $I=(-\tau)^{-n}$ one can solve Eq. (\ref{guage}) and obtain
\begin{equation}
\tilde{A} = \frac{1}{\sqrt{k}}\left( G_{-n-1}(\xi,-k\tau)+iF_{-n-1}(\xi,-k\tau)\right) \,,
\end{equation}
where $G$ and $F$ are Coulomb wave functions and $\xi  = \lambda n\gamma $. The superhorizon limit of this solution (i.e. for $-k\tau \ll \frac{1}{|\xi|} $) reads 
\begin{equation}
\tilde{A}(k,\tau) = -\sqrt{\frac{-\tau}{2\pi}} e^{\pi \xi} \Gamma(\abs{2n+1})\abs{ 2 \xi k \tau}^{-\abs{n+\frac{1}{2}}} \ .\label{gaugefield}
\end{equation}
Note that from $\xi>1$ one obtains exponential enhancement of sourced fluctuations. This superhorizon limit is exponentially enhanced compared to the subhorizon contribution, which is why we can neglect it in our calculations, see Appendix \ref{app:subh}. Moreover, only the polarization with positive $\xi$ can give us exponential magnification of second order perturbations, which is the mechanism that allows them to dominate over vacuum fluctuations. As a result the tensor spectrum will be chiral, contrary to the standard inflationary predictions. For the mode function given in \eqref{gaugefield}, the time dependence of the source term in Eqs. \eqref{eq:scsource},\eqref{eq:tensource} is given by $\frac{1}{I^2}$ \cite{r1} and can be taken outside the momentum integrals in Eqs. \eqref{eq:factorizationt} and \eqref{eq:factorizations}. This leads to factorization of momentum and time integrals.  

Once we have established that the momentum and time integrals factorize, the general form of perturbations $\hat X=\hat \zeta, \hat h$ are given by
\begin{equation}
\hat{X} = \mathcal{N}^{X} \mathcal{I}^X e^{2\pi \xi} \xi^{-|2n+1|} k^{3-|2n+1|} \hat{f}^X
\end{equation}
Thus, both scalar and tensor spectra can be written schematically in the following way for $\hat X=\hat \zeta, \hat h$ :
\bea
\langle X_{k}X_{k'} \rangle=\frac{2\pi^{2}}{k^3}\delta(\vec{k}+\vec{k}')({\cal P}^v_X(k)+{\cal P}^s_X(k)).\\
\mathcal{P}^s_{X}=\frac{2{\mathcal N}^{X\, 2} \mathcal{I}^{X\,2}}{2\pi^2}e^{4\pi \xi}\xi^{-2|2n+1|}k^{6-2|2n+1|}\times \abs{\hat{f^X(q)}}^2  \, \label{eq:pts}
\eea
where the numerical factor $\mathcal{N}^T= \mathcal{N}^S =\mathcal{N} $ is given by
\begin{equation}
    \mathcal{N} = \frac{2}{\pi} 4^n (n+1)^2 \Gamma(-2n-1).
\end{equation}
and $\mathcal{I}^X$ is defined by
\begin{equation}
\mathcal{I}^T = \int^{\tau} d\tau' \frac{G^T_k(\tau,\tau')}{a(\tau)a(\tau')} I^{-2} \, ,
\label{eq:timei}
\end{equation}
and
\begin{equation}
\mathcal{I}^S = \frac{1}{\sqrt{2\epsilon(\tau)}} \int^{\tau} d\tau'  \frac{\sqrt{2\epsilon(\tau')}\,C(\tau') G^S_k(\tau,\tau')}{a(\tau)a(\tau')} I^{-2}.
\label{eq:timeis}
\end{equation}
$\hat{f}^T,\hat{f}^S$ are the momentum integrals. 
\begin{equation}
\abs{\hat{f}^S}^2=\int \frac{d^3p}{(2\pi)^\frac{3}{2}}  \abs{\epsilon_i\epsilon_i}^2 \left(p\abs{\vec{k}-\vec{p}}\right)^{2n+1}\\
\end{equation}
and
\begin{equation}
\abs{\hat{f}^T}^2 = \int \frac{d^3p}{(2\pi)^\frac{3}{2}} \abs{ P_{\lambda}}^2 \left(p\abs{\vec{k}-\vec{p}}\right)^{2n+1}\\.
\end{equation}
where $P_{\lambda}\left(\vec{k},\vec{p},\vec{k}-\vec{p}\right)=\sum_{\lambda} \epsilon_i^{\lambda*}(\vec{k}) \epsilon_j^{\lambda*}(\vec{k})
\epsilon_i^{\lambda'}(\vec{p})\epsilon_j^{\lambda'}$. Notice that \eqref{eq:pts} guarantees $n_T=n_S-1=6-2|2n+1|$. So both spectra have the same tilt. For a viable scalar tilt, $n_S\simeq0.96$ we get $n=-2.01$ or $n=1.01$. Calculating $r$, we obtain
\begin{equation}
    r =  \frac{\mathcal{P}_T^s}{\mathcal{P}_S^s} = \frac{(\mathcal{N}^T \mathcal{I}^T)^2 \abs{\hat{f}^T}^2}{(\mathcal{N}^S \mathcal{I}^S)^2 \abs{\hat{f}^S}^2}
    \label{eq:newr1}
\end{equation}
Recall that $\mathcal{N}^T=\mathcal{N}^S$. Following calculations in \cite{r3}    $ \abs{\hat{f}^S}^2 \simeq \abs{\hat{f}^T}^2(1+\mathcal{O}(n_S-1))$. Hence, the contribution of the momentum integral to the power spectrum of perturbations are approximately the same, the only relevant difference between tensor and scalar perturbations for a scale invariant sourced spectrum is in the time integrals $\mathcal{I}^T$ and $\mathcal{I}^S$. Eq. \eqref{eq:newr1} can hence be simplified as
\be
r=\left(\frac{\mathcal{I}^{T}}{\mathcal{I}^{S}}\right)^2\, .
\label{eq:rs}
\ee
We will now show that for fast roll solutions $r$ has a lower bound that is ruled out by observations. As discussed in section \ref{sec:pertun} for slow contraction $\epsilon(\tau)>3$ and is monotonically increasing with $\tau$ (decreasing with $\abs{\tau}$), which imply that $\mathcal{I}^S$ obeys 
\begin{equation}
\abs{\mathcal{I}_S} \leq \abs{\int \frac{C(\tau') G_k^S(\tau,\tau')}{a(\tau) a(\tau')} I^{-2}} \, .
\label{eq:ub}
\end{equation}
We call this upper bound $\mathcal{I}^S_{m}$. Within the fast-roll regime one obtains $\epsilon$ very slowly varying with $\tau$, which gives $\mathcal{I}^S_{m} \simeq \mathcal{I}^S$. It follows from \eqref{eq:ub} and \eqref{eq:rs} that 
\begin{equation}
r \geq \left( \frac{\mathcal{I}^T}{\mathcal{I}^S_m} \right)^2 \, .
\label{eq:rsn}
\end{equation}
$\mathcal{I}^S_{m}$ and $\mathcal{I}^T$ differ by a factor of $C(\tau)$ and their respective Green's functions, which depend on vacuum fluctuations. As noted in the Sec. \ref{sec:scaleinv}, the scale invariance of the power spectrum requires $I=(-\tau)^{-n}$. Substituting this in Eq. \eqref{eq:C} 
\begin{equation}
C(\tau) =  1-2n \frac{\mathcal{H}}{\tau \phi'^2}=1-\frac{n}{\tau \epsilon \mathcal{H}} .
\end{equation}
One can express $C(\tau)$ in terms of $f_i$ parameters \eqref{eq:fi}, which gives
\begin{equation}
 C(\tau) =  1- \frac{n}{f_1+1-f_2} \, .
 \label{eq:ctau}
\end{equation}
$r$ being small requires $C(\tau)$ to be large. Within the fast-roll approximation one finds $\epsilon_N \ll \epsilon \Rightarrow f_1, f_2 \ll 1$. In such models $ C(\tau) =  1- \frac{n}{f_1+1-f_2}$ is approximately a constant and can be taken outside the time integral,
\begin{equation}
\mathcal{I}^S_m \simeq C(\tau) \int \frac{G_k^S(\tau,\tau')}{a(\tau) a(\tau')} I^{-2}
\label{eq:isfastroll}
\end{equation}
Substituting Eq.\eqref{eq:isfastroll} and Eq.\eqref{eq:gffastroll} in \eqref{eq:rsn} we obtain 
\begin{equation}
    r  \geq \left(\frac{\int \frac{G_k^S(\tau,\tau')}{a(\tau) a(\tau')} I^{-2}}{ C(\tau) \int \frac{G_k^S(\tau,\tau')}{a(\tau) a(\tau')} I^{-2} } \right)^2= \frac{1}{C^2}
\end{equation}
Using Eq.\eqref{eq:ctau} 
\begin{equation}
r \geq \frac{1}{C(\tau)^2} \simeq \left(1- \frac{n}{f_1+1-f_2} \right)^{-2} = \frac{1}{(1-n)^2}+\mathcal{O}(f_1^2,f_2^2).
\end{equation}
For $n=-2.01$ we get the desired red tilted $n_s=0.96$. Hence, in the fast-roll regime we obtain: \footnote{The $n=1$ branch gives roughly the same result. Once we use the duality of $n\rightarrow -1-n$, we get $r_{fast-roll}\simeq 1/(2+n)^2=1/9$ for $n=1$.}
\begin{equation}
r_{fast-roll} \simeq \frac{1}{9} \, ,
\end{equation}
where $r=1/9$ corresponds to $\epsilon = const$. This result is valid both for ekpyrosis and Intermediate contraction, as in both cases we have fast-roll with $f_1,f_2 \ll 1.$ The difference being that in ekpyrosis $f_2=0$, which is not the case in Intermediate contraction.\footnote{In \cite{r1} $r$ was calculated for ekpyrosis, and in a previous version it was claimed that $r\simeq 0.85$. This result was based on a small calculation error in expressing $I(\phi)$. The authors of \cite{r1} agree with our result and have published an erratum.}

One can imagine a situation, in which the $C(\tau)\simeq const.$ approximation is broken for few e-folds, while the Universe still remains in the slow contraction phase. This mechanism could be used to obtain viable value of $r$ for limited range of $N$. Assuming that $G_k^S \simeq G_k^T$ (which will give smallest possible r for fast-roll) one can set an upper bound on the tensor-to-scalar ratio. Consider the time integral $\mathcal{I}^S_m$ over a region $\Omega$
\begin{equation}
\int_{\Omega}  d\tau' C(\tau') \frac{G_k(\tau,\tau')}{a(\tau) a(\tau')} I^{-2}  \leq max(C(\tau)) \int_{\Omega}  d\tau' \frac{G_k(\tau,\tau')}{a(\tau) a(\tau')} I^{-2} \,,
\label{eq:zetalowerbound}
\end{equation}
where $max(C(\tau))$ is maximum value of $C(\tau)$ in the region $\Omega$. Combining Eqs. \eqref{eq:rs} and \eqref{eq:zetalowerbound} 
\begin{equation}
r \geq \left(\frac{1}{max(C(\tau))}\right)^2
\end{equation}
$C^{-2}$ is the lower bound on $r$. Hence, assuming $n=-2$ and $C^{-2} < r_{\max} \simeq 0.06$ gives 
\begin{equation}
C= 1+\frac{2}{f_1+ 1-f_2} > \frac{10}{\sqrt{6}}\, . \label{inr}
\end{equation}
Eq. \eqref{inr} combined with inequality \eqref{epsilon} limit the maximum number of e-folds of slow contraction for which $r$ is viable to be \footnote{see appendix \ref{app:appr} for details.}
\begin{equation}
\Delta N_{viable-r} \lesssim \log(2.19) \simeq 0.78 \, .
\end{equation}
Thus, regardless of the particular form of slow contraction, one can never generate viable values of $r$ for more than a fraction of an e-fold within the fast-roll approximation. This makes fast-roll slowly contracting models with perturbations sourced by a vector field generically inconsistent with the data. 


\section{Example of fast-roll: Intermediate contraction}\label{sec:ic}

\subsection{Unsourced fluctuations in Intermediate contraction}

The simplest and most discussed model of slow contraction is the power-law evolution of ekpyrosis \cite{BATTEFELD20151}. As mentioned in Sec. \ref{sec:slow} the Intermediate contraction \eqref{eq:intermediateV} is another good example of a theory, which is an exact solution of the Friedmann equations and fully satisfies the fast-roll approximation. In this section we explicitly calculate observables in the Intermediate contraction scenario, defined as
\be
a = e^{A(-t)^f},\quad V=\frac{8A^2}{(\beta+4)^2} \left(\frac{\phi}{\sqrt{2A\beta}}\right)^{-\beta}\left(6-\frac{\beta^2}{\phi^2}\right)
\ee
and $\beta = 4(1/f-1)$. For Intermediate contraction one obtains analytical solutions for the superhorizon modes. From \eqref{eq:intermediateV} and \eqref{eq:hsuper} one finds 
\begin{eqnarray}
h_0 = c_1-c_2\, \gamma \left(\frac{1}{f},3 A (-t)^f\right)=c_1-c_2\, \gamma \left(\frac{1}{f},3 N \right)\, , \label{eq:h0} \\
\zeta_0 = \tilde{c}_1-\tilde{c}_2\, \gamma \left(\frac{2}{f},3 A (-t)^f\right) =
\tilde{c}_1-\tilde{c}_2\, \gamma \left(\frac{2}{f},3 N\right) \, ,
\end{eqnarray} 
where  $c_i = c_i(k)$, $\tilde{c}=\tilde{c}_i(k)$ and $\gamma$ is the lower incomplete gamma function. Clearly one obtains a smooth freeze-out since $\gamma(1/f,3N) \to 0$ for $N \to 0$. 

Furthermore, in the case of Intermediate contraction it is easy to show that $v_S\simeq v_T$, since
\begin{equation}
\left(\frac{z''}{z} - \frac{a''}{a}\right)\frac{a}{a''} = \frac{f+2-6fN}{4N(2fN+f-1)} \simeq -\frac{1}{2N} \ll 1 \, .
\end{equation}
In the limit $fN\ll1$ one finds $\frac{a''}{a} \simeq -\frac{b}{\tau^2}$, where $b = f N$. $b$ is extremely slowly varying with respect to $\tau$ and constant $b=f N_*$ is a reasonable approximation.
\begin{figure}[H]
\centering
\includegraphics[width=7cm]{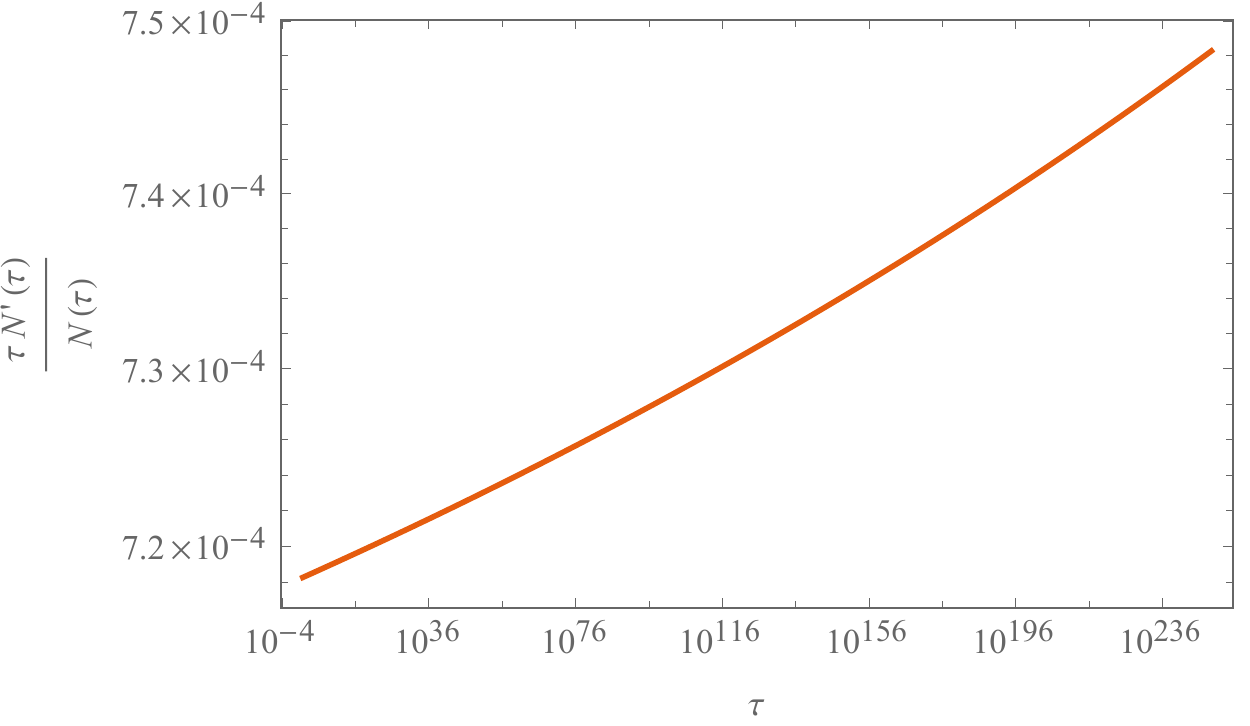}
\caption{The evolution of $\tau \frac{N'(\tau)}{N(\tau)}$ for the last 60 e-folds of contraction, where $N=60$ corresponds to big $\tau$. We have assumed $A=100$ and $f=1/1500$. One can see that $N$ very weakly depends on $\tau$ and therefore during the last 60 e-folds we can assume that $N$ as a function of $\tau$ is approximately constant.}
\label{fig:nuvariation}
\end{figure}
The solution of \eqref{eq:MSeq} is
\begin{equation}
v_S \simeq v_T \simeq \frac{\sqrt{\pi (-\tau)}}{2} H_{\frac{1}{2} - b}^{(1)}(-k\tau) \, ,
\label{eq:vacsol}
\end{equation}
where $H_{\frac{1}{2} - b}^{(1)}$ is a Hankel function. In the superhorizon limit one finds
\begin{equation}
\mathcal{P}_S = \frac{16}{\pi^2 b^2 M_{pl}^2 } e^{-2A} \abs{\Gamma\left(\frac{1}{2}-b\right)}^2 k^{2+2b} \, .
\label{eq:vactensor}
\end{equation}
 The spectrum obtains it's pivot scale value at $k=k_\star = a_\star H_\star$, where $a_\star=a(N_\star)$ and $H_\star=H(N_\star)$. Clearly $n_T=n_S-1= 2+2b\simeq 2$, so the scalar spectrum is highly blue tilted, which is inconsistent with the data \cite{Aghanim:2018eyx}. Thus, it is necessary to involve extra fields in order to generate a viable spectrum.

\subsection{Sourced fluctuations in Intermediate contraction}

\begin{figure}[H]
\centering
\includegraphics[width=0.45\textwidth]{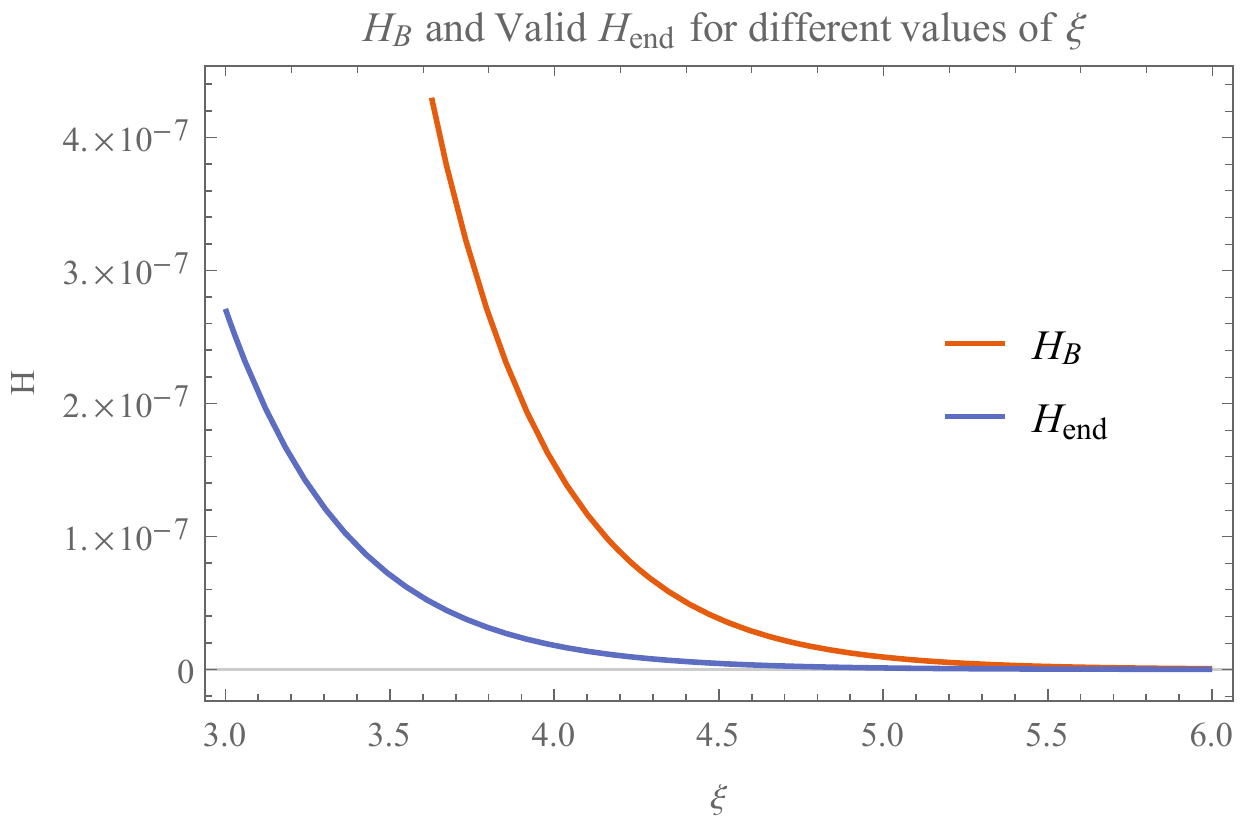} 
\hspace{0.5cm}
\includegraphics[width=0.45\textwidth]{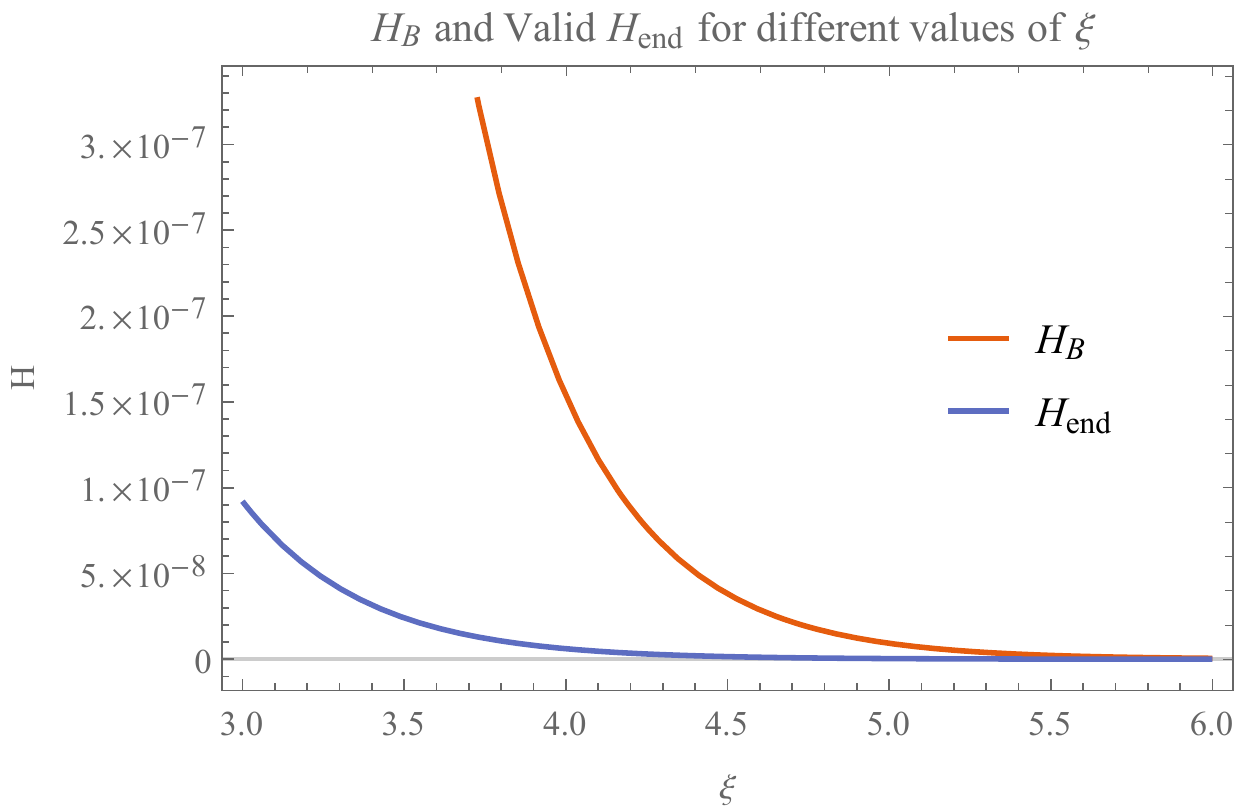} 
\caption{Values of $|H_{end}|$ and $|H_B|$  as a function of $\xi$ for the scalar spectrum normalized to observable values and $b=\frac{1}{20}$. Left and right panels correspond to $A=60$ and $A=62$ respectively. Note how even small changes in $A$ lead to significant changes in $H_{end}$ and $H_B$.}
\label{fig:amplitude}
\end{figure}

The energy density of the gauge field redshifts like $\rho_A \propto \tau^{-4}$, while $\rho_{\phi} \propto \tau^{-2}$. As a result the gauge field will dominate the evolution of the Universe at some time of backreaction $\tau = \tau_B$ defined by $\rho_A(\tau_B) = \rho_\phi(\tau_B)$ and invalidate the analysis. Hence, for consistency, the sourced fluctuations must shut down before we reach the backreaction limit.  We conduct calculations similar to \cite{r1} and conclude that backreaction limit for Intermediate contraction (when $n$ is close to $-2$) is given by
\begin{equation}
    \frac{H_B}{M_{pl}} = \sqrt{\frac{48 \pi^2 (n+2)}{\mathcal{N}}}\frac{b^2}{(1-b)^2} \xi^{\frac{
     3}{2}} e^{-\pi \xi}
\end{equation}
where $H_B$ is Hubble at backreaction. For Intermediate contraction, it is possible to ignore the backreaction at early epochs since the gauge field does not contribute to the spectrum at this time. This happens during the accelerated contraction phase, where the potential is positive $V(\phi)>0$. Thus, there is some finite $\tau$ beyond which backreaction can be ignored. As a result there is no IR divergence for $n<-2$ in this model, unlike ekpyrosis \cite{r1}. Upon choosing $n=-2.01$
\begin{equation}
\mathcal{P}_S \propto k^{8+4n} \propto k^{-0.04}
\end{equation}
and $n_s = 0.96$ in agreement with observations \cite{Aghanim:2018eyx}. We use approximate expressions $a\simeq (-\tau)^b$, $ G_k(\tau,\tau') \simeq \Theta(\tau-\tau') \tau' $ in order to evaluate the time integral $\mathcal{I}^S$  and evaluate Eq.(\ref{eq:pts}) to obtain the power spectrum for Intermediate contraction. Substituting numerical values of all the constants and $\abs{\hat{f}^S}^2$ from \cite{r3} we obtain
\begin{equation}
P^s_S  = \frac{11.1}{256 r \pi^6 (1-n_s)} \frac{e^{4\pi \xi}}{b^4 \xi^6} \left(\frac{H_{end}}{M_{pl}}\right)^4 \left(\frac{k}{H_{end}}\right)^{n_s-1}.
\label{eq:icspectrum}
\end{equation}
The measured amplitude of the power spectrum is $2.1 \times 10^{-9}$ \cite{Aghanim:2018eyx}. By choosing appropriate values of parameters $\xi,b,A$ and $H_{end}$ it is possible to match power the spectrum in \eqref{eq:icspectrum} with observations, while fulfilling the backreaction bound. In figure \ref{fig:amplitude} we show a range of values of parameters $H_{end}$ and $\xi$ for which one can obtain the correct amplitude given some value of $A$ and $b$. One can observe that $H_{B}\gg H_{end}$ especially for larger values of $A$. 
Using the definition of $f_1,f_2$ from Eq.\eqref{eq:fi} as well as $\tau(N)$ relation from \eqref{conf} one finds
\begin{equation}
f_1 = \frac{fN(1 - fN)}{1 + f N (-2 + f + f N)} \simeq fN \; \qquad \; f_2 = \frac{f - 2 f^2 N - (f-1) f^3 N^2}{(1 + f N (-2 + f + f N))^2} \simeq  f.
\end{equation}
Clearly $f_1 \simeq fN \ll 1$ and $f_2 \simeq f \ll 1$ as we expect for fast-roll solutions. In addition $f_2$ and $f_1$ are approximately constant. Thus, the case of Intermediate contraction is a perfect example of the fast-roll analysis discussed in Sec. \ref{sec:Green'sfi}, and $C(\tau)$ is approximately a constant resulting in
\begin{equation}
r \simeq \frac{1}{C(\tau)^2} \simeq \left(\frac{1}{1-n+n(fN-f)}\right)^2 \simeq \frac{1}{9}+\mathcal{O}((fN)^2) \,,
\end{equation}
for $n=-2.01$ as required from the observed $n_S\simeq0.96$. For any realistic Intermediate contraction $r\simeq1/9$, which is consistent with our general analysis in Sec. \ref{sec:r}. 


\section{Conclusions} \label{sec:concl}

 We have discussed sourced fluctuations in the framework of a slowly contracting Universe described by a single scalar field, the bouncer field. Our implicit assumption is that the spectra derived here are not modified by the bounce. In the quest for a model that is in accord with the CMB data one can roughly consider different sources, different slowly contracting backgrounds, and different couplings between the source and the bouncer field. We have analyzed the latter two. For the given $U(1)$ source at hand, only a specific coupling, $I \propto \tau^2$ or $I\propto \tau^{-1}$ gives a viable nearly scale invariant scalar spectrum. We derived rather general slowly contracting backgrounds that are not necessarily ekpyrosis. For the different slowly contracting backgrounds, we have shown that in the fast-roll approximation defined as $|\epsilon_N/\epsilon|\ll 1$ the vacuum fluctuations and as a result the Green's functions for scalar and tensors are fixed to be the same up to the standard $\epsilon$ normalizations and a time-dependent function, $C(\tau)$. The time dependent function is approximately constant in the fast-roll approximation. As a result, for a given background we get $r\simeq 1/9$, which is inconsistent with current data. A temporary deviation from $C(\tau)\simeq const$ allows for a viable $r$ only during a single e-fold, which is still inconsistent.
Nevertheless, the sourced fluctuations discussed here could be at work on Laser Interferometer scales, provided that the CMB observations are explained by some other mechanisms. In such a case, the expected signal is that of a \textit{chiral} stochastic background of gravitational waves with a blue-tilt $0.15\lesssim n_T\lesssim 2.7$, as discussed in \cite{r3}.

Let us discuss a few possible loopholes in our no-go theorem starting with the fast-roll approximation. Consider a slowly contracting background that is not fast-roll. In such a case, there are several places where the no-go fails. First, the vacuum fluctuations of the tensors can be different from the scalar ones, while both freeze-out outside the horizon. As a result we shall obtain different spectra already at the vacuum level. Second, due to the deviation from fast-roll, the Green's functions of scalars and tensors will be modified as well yielding a different $r$. Finally, the time dependent function $C(\tau)$ is significantly modified once we deviate from fast-roll. Any one of these three loopholes can modify the outcome  enough to yield a viable model. 

The second obvious possibility is considering a different source. Here the crucial aspect is whether $J_S\simeq J_T$, or not. While it is difficult to make general statements, it seems we can point on potential promising avenues. Considering the coupling to be of the form $I^2(\phi)\mathcal{L}_{source}$, then at leading order $J_S\sim \mathcal{L}_{source}$. Considering the coupling to tensors, in the traceless transverse part of the energy momentum tensor, we also have $J_T \supset \mathcal{L}_{source}$, but with additional contributions. Thus if for both $J_{S,T}\sim \mathcal{L}_{source}$ then it seems we will again reach a null result, while a deviation from the similarity of sources could yield a viable $r$ and scalar spectrum.


\appendix

\textbf{Acknowledgements:}
The work was partly funded by
project RA1800000062.

\section{Comparison of tensor and scalar Green's functions} \label{app:Green'scomp}
Within the fast-roll regime in \eqref{eq:appzpp} one finds
\begin{equation}
\frac{z''}{z} \simeq \frac{a''}{a} -\frac{ \epsilon_N}{2} \mathcal{H}^2.
\end{equation}
where $\epsilon_N<0$. Let us define the following operator 
\begin{equation}
\Delta_i = \dfrac{d^2}{d\tau^2} + \left(k^2 - \frac{d_i''}{d_i}\right)
\end{equation}
with Bunch-Davies vacuum boundary conditions. 
Then
\begin{equation}
\Delta_T-\Delta_S = \frac{z''}{z}-\frac{a''}{a} \simeq -\frac{ \epsilon_N}{2} \mathcal{H}^2.
\label{op}
\end{equation}
Green's functions are inverse of $\Delta$ operators. From Eq. \eqref{op} we obtain
\begin{equation}
\Delta_T G_S = \Delta_S G_S+\frac{ \epsilon_N}{2} \mathcal{H}^2 G_S =   \delta(\tau-\tau')+\frac{ \epsilon_N}{2} \mathcal{H}^2 G_S
\end{equation}
whose solution is given by
\begin{equation}
G_S = G_T+\int \frac{ \epsilon_N}{2} \mathcal{H}^2 G_S G_T .
\end{equation}
This is a series solution in $G_T$, i.e. the Dyson series expansion of the Green's function. For fast-roll the $\epsilon_N$ correction is very small and we can approximate the series at first order which gives
\begin{equation}
G_T-G_S = -\int \frac{ \epsilon_N}{2} \mathcal{H}^2  G_T G_T.
\end{equation}
Since $\epsilon_N$ is negative, $-\int \frac{ \epsilon_N}{2} \mathcal{H}^2  G_T G_T \geq 0 $. Thus
\begin{equation}
G_T \geq G_S
\end{equation}
One finds $G_T=G_S$ for $\frac{z''}{z}=\frac{a''}{a}$. Any deviation from the exact equality increases
$r$, and $r$ will be minimized for $G_T = G_S$.

\section{Conditions for viable r}\label{app:appr}
From Eq. \eqref{inr} we conclude that in order to get viable $r$, the $f_1$ parameter must satisfy
\begin{equation}
f_2 = f_{1,N} \gtrsim \frac{1}{3}+f_1 \, . \label{eq:f2good}
\end{equation}
We split the integral (\ref{eq:zetalowerbound}) over the domain $\Omega$ into two regions $\Omega(f_2<\frac{1}{3}+f_1)$ and $\Omega(f_2>\frac{1}{3}+f_1)$
\begin{equation}
\begin{split}
\int_{\tau_{end}}^{\tau} d\tau' C(\tau') \frac{G_k(\tau,\tau')}{a(\tau) a(\tau')}  I^{-2} &= \int_{\Omega(f_2<\frac{1}{3}+f_1)}  d\tau' C(\tau') \frac{G_k(\tau,\tau')}{a(\tau) a(\tau')}  I^{-2} \\&
\;\;\;+ \int_{\Omega(f_2>\frac{1}{3}+f_1)} d\tau' C(\tau') \frac{G_k(\tau,\tau')}{a(\tau) a(\tau')} I^{-2}
\end{split}
\end{equation}
\begin{equation}
\begin{split}
\int_{\tau_{end}}^{\tau} d\tau'& C(\tau') \frac{G_k(\tau,\tau')}{a(\tau) a(\tau')} I^{-2} \\
&< max( C(\tau))_{\Omega(f_2<\frac{1}{3}+f_1)}  \int d\tau \frac{G_k(\tau,\tau')}{a(\tau) a(\tau')} I^{-2} \;\;\; \tau  \in \; \Omega(f_2<\frac{1}{3}+f_1) \\
&<  max(C(\tau))_{\Omega(f_2>\frac{1}{3}+f_1)}  \int d\tau' \frac{G_k(\tau,\tau')}{a(\tau) a(\tau')}  I^{-2} \;\;\;  \tau  \in \; \Omega(f_2>\frac{1}{3}+f_1).
\end{split}
\end{equation}
Tensor to scalar ratio r is viable only in the region $\Omega(f_2>\frac{1}{3}+f_1)$. Any contribution from the second region will render $r$ inconsistent with the data. Using Gronwall's inequality in \eqref{eq:f2good} one finds
\begin{equation}
f_1> -\frac{1}{3}+ e^{\Delta N}\left(f_1(N_0)+\frac{1}{3}\right) \, .
\end{equation}
where $f_1(N_0)$ is the value of $f_1$ at $N=N_0$ and $\Delta N$ is number of e-folds of $\Omega_{f_2>\frac{1}{3}+f_1}$. This inequality is valid for all $N_0<N$. Let $N_0$ be the smallest $N \; \in \; \Omega_{f_2>\frac{1}{3}+f_1} $. We do not know the exact value of $f_1(N_0)$, but for the sake of our analysis it is enough to note that $f_1(N_0)>0$.\\
Obtaining $C^{-2}<r_{\max}$ requires $f_1$ to simultaneously satisfy two differential inequalities, \eqref{epsilon} and \eqref{eq:f2good}. Solving inequality (\ref{epsilon}) we obtain that
\begin{equation}
f_1 < \frac{1}{2}+\left(f_1(N_0)-\frac{1}{2}\right) e^{-2 \Delta N} \, .
\end{equation}
Both inequalities \eqref{epsilon} and \eqref{eq:f2good} can only be satisfied at the same time if 
\begin{equation}
e^{3 \Delta N} \left( f_1(N_0)+\frac{1}{3} \right) < \frac{5}{6} e^{2\Delta N} +  \left( f_1(N_0)-\frac{1}{2} \right)\, . \label{eq:xinequal}
\end{equation}
The \eqref{eq:xinequal} inequality has the following solution
\begin{equation}
0<f_1(N_0)<\frac{2}{9} \quad \text{and} \quad  1< e^{\Delta N} <\frac{5-6a}{12a}+ \frac{1}{12} \sqrt{\frac{25+60 a-108 a^2}{a^2}}
\label{fff}\, ,
\end{equation}
where $a = f_1(N_0)+\frac{1}{3}>\frac{1}{3}$. For $f_1(N_0)$ constrained by equation (\ref{fff}), the constraint on $\Delta N$ is given by
\begin{equation}
1<e^{\Delta N}<2.19
\end{equation}

\section{Conformal time for Intermediate contraction} \label{app:Appconf}

Consider the conformal time in an Intermediate contracting Universe
\begin{equation}
\tau = \int_{-t}^0 \frac{1}{a(t)} dt = \frac{A^{-\frac{1}{f}}}{f}\gamma \left(\frac{1}{f},A (-t)^f\right) = \frac{A^{-\frac{1}{f}}}{f}\gamma \left(\frac{1}{f},N \right) \, .
\end{equation}
Using Kummer's confluent hypergeometric function $M(1,s+1,x)$ we can arrive at an approximate expression for $\gamma(s,x)$. It is a well known result that
\begin{equation}
\gamma(s,x) = \frac{x^s e^{-x}}{s} M(1,s+1,x).
\end{equation}
Series expansion for $M(1,s+1,x)$ is
\begin{equation}
M(1,s+1,x) = 1+\sum_{k=1}^{\infty} \frac{x^k}{\Pi_k (s+k)}.
\end{equation}
Clearly when $\frac{x}{s}\ll 1$ this series can be approximated by the geometric series $\sum_k \left(\frac{x}{s}\right)^k$. We may estimate an upper bound for error in this approximation by subtracting the two series
\begin{equation}
\Delta = \sum_k \frac{x^k}{s^k}-\left( 1+\sum_{k=1}^{\infty} \frac{x^k}{\Pi_k (s+k)} \right)\leq \sum_{k=1}^{\infty} \frac{x^k}{s^{k+1}}= \frac{x}{s^2(1-\frac{x}{s})}
\end{equation}
For sufficient number of e-folds of slow contraction, we have already shown that $\frac{1}{f}\geq 3N_{max}$ is needed. Thus $\frac{x}{s}= f N \leq \frac{1}{3}. $ Hence it is possible to employ the series summation outlined above. Writing $\tau$ in terms of $N$ gives:
\begin{equation}
\tau = \frac{A^{-\frac{1}{f}}}{f}\; \gamma\left(\frac{1}{f},N \right) \approx A^{-\frac{1}{f}}\; N^{\frac{1}{f}} e^{-N} \frac{1}{1-f N} \, .
\label{conf}
\end{equation}
When $fN = \frac{1}{3}$ we see that the upper bound for error in our approximation is $\frac{f^2 N}{1- fN} = \frac{f}{2}$. It is even smaller for smaller $fN$. Such error is incredibly small for the values of $f$ that are of interest. 

\section{Subhorizon contributions} \label{app:subh}
In this appendix we show that it is in general possible to ignore the subhorizon contributions to the sourced spectrum. To see this note that gauge fields that source perturbations are solutions to Eq.\eqref{guage}. Upon a change of variables from $(-\tau)$ to $x=-k\tau$, Eq.\eqref{guage} takes the form:
\begin{equation}
\dfrac{d^2\tilde{A}_{\lambda}}{dx^2}+\left(1 -2\lambda \gamma  \frac{1}{I}\dfrac{dI}{dx} - \frac{1}{I} \dfrac{d^2I}{dx^2}\right) \tilde{A}_{\lambda} = 0 \, , \label{guagex}
\end{equation}
with $A=\frac{\tilde{A}(x)}{I(x)}$. Outside the horizon $I_{xx}/I\gg1$, leading to superhorizon results that we explain in the rest of the paper. 
Let us look at the subhorizon solution for this equation, inside the horizon $I_{xx}/I$, $I_x/I\ll1$ leading to Minkowski solutions, i.e.
\begin{equation}
    \tilde{A} \simeq \frac{1}{\sqrt{2k}} e^{\pm i k \tau}
\end{equation}
which are independent of $I$. The source term for a $U(1)$ gauge field
\begin{equation}
    J_{\lambda}(\tau,k)  \propto I^2 \int  d^3x  (\hat{E_i} \hat{E_i} + \hat{B_i} \hat{B_i} + 2\gamma \hat{B_i} \hat{E_i}), \,\, \\
\end{equation}
where 
\begin{equation}
\begin{split}
    \hat{E_i} &= \partial_{\tau} \frac{ \tilde{A}}{I} = \left(  \frac{1}{I} + \frac{I^{'}}{k I^2} \right)\sqrt{\frac{k}{2}} e^{i k x} \hat{\epsilon}_i \left[ \hat{a}(\vec{k})+\hat{a}^{\dagger}(-\vec{k})\right]  \\
    & \simeq   \frac{1}{I} \sqrt{\frac{k}{2}} e^{i k x} \hat{\epsilon}_i \left[ \hat{a}(\vec{k})+\hat{a}^{\dagger}(-\vec{k})\right] 
    \end{split}
\end{equation}
and  
\begin{equation}
    \hat{B_i} = \frac{1}{I}\sqrt{\frac{ k}{2}} \hat{\epsilon}_i e^{i k x} \left[ \hat{a}(\vec{k})+\hat{a}^{\dagger}(-\vec{k})\right]. \
\end{equation}
Sourced scalar perturbations are of the form given in Eq.\eqref{eq:vi}. One can extract the k dependence of perturbations sourced by subhorizon contributions from the integral in Eq.\eqref{eq:vi} where Green's functions are of the form given in Eq.\eqref{eq:greenfss} assuming a fast-rolling back ground. Ignoring all the k independent factors, scalar perturbations are given by \\\
\begin{equation}
    \begin{split}
     &\hat{\zeta}  = \int G(\tau,\tau') J d\tau' \propto \int \frac{G(\tau,\tau') C(\tau')}{a(\tau')}  I^2 d\tau' \int d^3 p \, (\hat{B_i} \hat{B_i} +  2\gamma \hat{B_i} \hat{E_i} +  \hat{E_i} \hat{E_i} ) \\
        &\propto \int \frac{C(\tau') G(\tau,\tau')}{a(\tau')}   d\tau' \int d^3 p   \sqrt{ p\abs{\vec{k}-\vec{p}}}
        \,\left[ \hat{a}(\vec{p})+\hat{a}^{\dagger}(-\vec{p})\right] \left[ \hat{a}(\vec{k}-\vec{p})+\hat{a}^{\dagger}(\vec{p}-\vec{k})\right].
    \end{split}
\end{equation} 
Changing variables from $(-\tau)$ to $x=-k \tau$ in the above integral and noting that for fast-roll, contributions from scale factor $a$ and $C(\tau)$ to the scale dependence are insignificant, one can derive that scalar perturbations are proportional to 
\begin{equation}
\begin{split}
\hat{\zeta}  &
\propto \int \frac{\sin(x)}{k^2}  dx' \int d^3 p   \sqrt{ p \abs{\vec{k}-\vec{p}}} 
 \left[ \hat{a}(\vec{p})+\hat{a}^{\dagger}(-\vec{p})\right] \left[ \hat{a}(\vec{k}-\vec{p})+\hat{a}^{\dagger}(\vec{p}-\vec{k})\right].
\end{split}
\end{equation}
As a result, the correlation function $ \langle \hat{\zeta} \hat{\zeta} \rangle$ is proportional to $k$, and power spectrum to $k^4$ and subhorizon contributions cannot give a scale invariant spectrum. Hence, it is enough to look for scale invariance within perturbations sourced by the superhorizon solutions to the gauge fields. In Sec. \ref{sec:scaleinv} we saw that superhorizon contributions produce a scale invariant spectrum if the coupling is of the form $I\propto(-\tau)^{-n}$. As explained in Sec. \ref{sec:r} superhorizon perturbations are exponentially enhanced. Thus, blue tilted spectrum of subhorizon solutions do not contribute to the scalar power spectrum significantly. Therefore, it is possible to ignore subhorizon contributions. \\
 \\
 Let us also examine the contribution of magnetic field to the correlation functions. From the previous discussion it is clear that the subhorizon scales cannot produce a scale invariant spectrum. For power law $I$ the subhorizon solutions are exponentially suppressed.  Superhorizon solutions for magnetic fields are of the form given by
 \begin{equation}
     \hat{B_i} = D_1(k) k_i + D_2(k) k_i \int \frac{1}{I(x)^2} dx.
 \end{equation}
Scale invariance require power law form of $I$, as the difference between the contribution of the magnetic or electric field is simply another power of $k$. For power law form of $I$ we know that  
for time dependent terms in $\vec{A}$,  $\hat{E_i}\sim x \hat{B_i}$, hence $\hat{E_i}\gg\hat{B_i}$ in the superhorizon limit. The contribution of the constant term is given by
\begin{equation}
\begin{split}
J_{\lambda}^S (\tau,\Vec{k})\simeq & \frac{I^2}{2 a}\int \frac{d^3p}{2\pi^3} 
\epsilon_i^{\lambda}(\vec{p})\epsilon_j^{\lambda}(\vec{p}-\vec{k}) \left(D_1(k)D_1(\abs{\Vec{k}-\vec{p}}) \right) k^2\\
&\;\;\;\;\;\;\;\; \times C(\tau) \sqrt{2\epsilon(\tau)} \left[ \hat{a}_{\lambda}(\vec{p})+\hat{a}^{\dagger}_{\lambda}(-\vec{p})\right]\left[ \hat{a}_{\lambda}(\vec{k}-\vec{p})+\hat{a}^{\dagger}_{\lambda}(-\vec{k}+\vec{p})\right].
\end{split}
\label{eq:scsourcegenm}
\end{equation}
Thus, contribution of the constant magnetic field to the scalar power spectrum is given by
\begin{equation}
    \mathcal{P}_{SB} \propto \left(k^4 \int_{x_{end}}^{1} x' dx'  \frac{ \sqrt{2\epsilon} C(x')}{a(x)} \frac{I^2}{k^2} \right)^2 .
\end{equation}
For $I=\tau^2$ we see that ${P}_{SB} \simeq (\xi x_{end})^8 P_S$, which is completely negligible.

Further validating our claims one can see that the largest contribution to the time integral is obtained from first order term in the expansion of the Green's function. At superhorizon scales, for fast-roll, the time integral in Eq.\eqref{eq:timeis} in $x'=-k\, \tau'$ is approximately, 
\begin{equation}
\mathcal{I}_S \propto k^{-2}\int \sin(x')  I^{-2}(x') dx'.
\end{equation}
Time dependence of the source $I$ is monotonic. Expanding in $x'$
\begin{equation}
 \mathcal{I}_S  \propto k^{-2} \int \sum_{i=0}^{\infty} (-1)^i \frac{(x')^{2i+1}}{(2i+1)!} \frac{1}{I(x')^2} dx' = \mathcal{I}(1/k)-\mathcal{I}(x_{end}).
\end{equation}
We are only interested in the dominant term of the expansion. If $I(x)$ is decreasing faster than $x^{-\frac{1}{2}}$, $I(x_{end})$ is the dominant term and the first order term is bigger than all the higher order terms, as this term has the highest negative power in $x_{end}$. Similarly one can show that the first term is dominant when $I$ is an increasing function. This is clear when we look at the time integral after we have chosen a form of $I$. For the realistic case of $I =(-x)^{-n}$ and $n = -2$ one finds 
\begin{equation}
\mathcal{I}_S \propto k^{-2} \sum_{i=0}^{\infty} (-1)^i\left( \frac{1}{(2i+1)!}-\frac{x_{end}^{2i-2}}{(2i+1)!(2i-2)} \right).
\end{equation}
Clearly the dominant term is $\mathcal{I}_S \propto k^{-2}x_{end}^{-2}$ for $i=0$. Hence we have shown that it is enough to consider only the first term in the expansion of the Green's function for fast roll while calculating power spectrum of sourced perturbations. 

\bibliographystyle{JHEP}
\bibliography{references}
 
\end{document}